\PassOptionsToPackage{pdfpagelabels=false}{hyperref} 
\documentclass[a4paper,fleqn,usenatbib]{mnras}

\usepackage[T1]{fontenc}
\usepackage{ae,aecompl}

\usepackage{graphicx}	
\usepackage{amsmath}	
\usepackage{amssymb}	
\usepackage{color}

\def\aj{AJ}
\def\apj{ApJ}
\def\aap{A\&A}
\def\apjl{ApJL}
\def\apjs{ApJS}
\def\aaps{A\&AS}
\def\mnras{MNRAS}
\def\araa{ARA\&A}

\def\pasp{PASP}

\def\pasj{PASJ}

\title[H$ \alpha $ Kinematics of S$ ^{4} $G Spiral Galaxies-III]
  {H$ \alpha $ Kinematics of S$ ^{4} $G Spiral Galaxies - III. Inner rotation curves}

\author[S. Erroz-Ferrer et al.]
{Santiago Erroz-Ferrer,$^{1,2,3}$\thanks{Email: serrozfe@phys.ethz.ch} Johan H. Knapen,$^{1,2}$ Ryan Leaman,$^{1,2,4}$   
\newauthor Sim\'on D\'iaz-Garc\'ia,$^{5,6}$ Heikki Salo,$^{5}$ Eija Laurikainen,$^{5,6}$ Miguel Querejeta,$^{4}$
\newauthor  Juan Carlos Mu\~noz-Mateos,$^{7}$ E. Athanassoula,$^{8}$  Albert Bosma,$^{8}$
\newauthor Sebastien Comer\'on,$^{5,6}$ Bruce G. Elmegreen$^{9}$ and Inma Mart\'inez-Valpuesta$^{1,2}$ \\     
$^{1}${Instituto de Astrof\'isica de Canarias, V\'ia L\'actea s/n 38205 La Laguna, Spain}\\
$^{2}${Departamento de Astrof\'isica, Universidad de La Laguna, 38206 La Laguna, Spain}\\
$^{3}${Department of Physics, Institute for Astronomy, ETH Zurich, CH-8093 Zurich, Switzerland}\\
$^{4}${Max-Planck-Institut f\"ur Astronomie / K\"onigstuhl 17 D-69117 Heidelberg, Germany}\\
$^{5}${Astronomy Division, Department of Physical Sciences, FIN-90014 University of Oulu, P.O. Box 3000, Oulu, Finland}\\
$^{6}${Finnish Centre of Astronomy with ESO (FINCA), University of Turku, V\"ais\"al\"antie 20, FI-21500, Piikki\"o, Finland}\\
$^{7}${European Southern Observatory, Casilla 19001, Santiago 19, Chile}\\
$^{8}${Aix Marseille Universit\'e CNRS, LAM (Laboratoire d'Astrophysique de Marseille) UMR 7326, 13388, Marseille, France}\\
$^{9}${IBM Research Division, T.J. Watson Research Center, Yorktown Hts., NY 10598, USA}}

\date{Accepted 2015 xxxx. Received 2015 xxx; in original form 2015 xxx}

\pubyear{2015}

\begin{document}
\label{firstpage}
\pagerange{\pageref{firstpage}--\pageref{lastpage}}
\maketitle

\begin{abstract}
We present a detailed study of the shape of the innermost part of the rotation curves of a sample of 29 nearby spiral galaxies, based on high angular and spectral resolution kinematic H$ \alpha $ Fabry-Perot observations. In particular, we quantify the steepness of the rotation curve by measuring its slope $d_{R}v_{\rm c}(0) $. We explore the relationship between the {inner} slope and several galaxy parameters, such as stellar mass, maximum rotational velocity, central surface brightness ($ \mu_{0} $), bar strength and bulge-to-total ratio. Even with our limited dynamical range, we find a trend for low-mass galaxies to exhibit shallower rotation curve {inner} slopes than high-mass galaxies, whereas steep {inner} slopes are found exclusively in high-mass galaxies. {This trend may arise from the relationship between the total stellar mass and the mass of the bulge, which are correlated among them.} We find a correlation between the {inner} slope of the rotation curve and the morphological \textit{T}-type, complementary to the scaling relation between $d_{R}v_{\rm c}(0) $  and $ \mu_{0} $ previously reported in the literature. Although we find that the {inner} slope increases with the Fourier amplitude $ A_{2} $ and decreases with the bar torque $ Q_{\rm b} $, this may arise from the presence of the bulge implicit in both $ A_{2} $ and $Q_{\rm b} $.  As previously noted in the literature, the more compact the mass in the central parts of a galaxy (more concretely, the presence of a bulge), the steeper the {inner} slopes. {We conclude that the baryonic matter dominates the dynamics in the central parts of our sample galaxies.}
\end{abstract}




\section{Introduction}
Rotation curves have been extensively used in the literature as tracers of the distribution of mass in spiral galaxies. The link between the shape of the rotation curve and the amount of luminous (baryonic) and dark matter (DM) provides important information about the formation and evolution of galaxies. One of the first pieces of evidence for the presence of DM in galaxies was the flat part of H{\sc i} rotation curves (\citealt{Bosma1981}; \citealt{vanAlbada1985}; \citealt{vanAlbada1986}; \citealt{Begeman1987}), because such flat H{\sc i} rotation curves extend to much larger radii than optical data, where DM is thought to dominate the gravitational potential (\citealt{Rogstad1972}; \citealt{Roberts1975}; \citealt{Bosma1978,Bosma1981}).



On the one hand, \citet{Burstein1985} and \citet{Rubin1985} defended that the shape of the rotation curve (and therefore the mass distribution) does not depend on morphological type, luminosity, total mass, bulge-to-disc ratio or other global properties of the galaxies, suggesting that the form of the gravitational potential is not correlated with the light distribution. \citet{Persic1991} and \citet{Persic1996} presented the idea of a universal rotation curve that only depends on the total luminosity of a galaxy. But on the other hand, other studies  defend that the total luminosity of the galaxy and other galaxy properties (i.e., presence of a bulge) influence the shape of the rotation curve (\citealt{Corradi1990}; \citealt{Casertano1991}; \citealt{Broeils1992}; \citealt{Verheijen1997,Verheijen2001}; \citealt{Swaters1999}; \citealt{Matthews2002}; \citealt{Swaters2003} and \citealt{Sancisi2004}, among others).

Several studies of H{\sc i} rotation curves have analysed the shape of the rotation curve (e.g., \citealt{deBlok1996}; \citealt{Swaters1999}; \citealt{Cote2000}; \citealt{Verheijen2001}; \citealt{Gentile2004}; \citealt{Noordermeer2007}; \citealt{Swaters2009, Swaters2011, Swaters2012}), including all types of morphologies, luminosities and galaxy properties {(e.g., both high- and low-surface brightness galaxies)}. Few studies have performed a quantitative study of the shapes of rotation curves. In particular, \citet{Swaters2009} parametrised the shape of the rotation curve with the logarithmic slope $ S= \Delta {\rm log}v / \Delta {\rm log}R$, looking for correlations between the logarithmic slope and the central surface brightness or morphological type. They concluded that for both spiral and late-type dwarf galaxies, the correlation between the light distribution and the inner rotation curve shape suggests that galaxies with higher central light concentration also have  higher central mass densities, which implies that the luminous mass dominates over the DM mass in the gravitational potential in the central regions. 

One of the most recent quantitative studies of the {inner} slopes of rotation curves was presented by \citeauthor{Lelli2013} (2013, LFV13 hereafter). They found a scaling relation between the circular velocity gradient $ d_{R}V(0) $ (what we will refer to as the slope of the circular velocity curve) and the central surface brightness $\mu_{0}$ over more than two orders of magnitude in $ d_{R}V(0) $ and four orders of magnitudes in $\mu_{0}$. This scaling relation for disc galaxies shows a clear relationship between the stellar density of a galaxy in the centre and the inner shape of the potential well, also for low-surface brightness galaxies where DM is thought to dominate in the central regions.


The {inner} shape of rotation curves derived from {low-resolution} H{\sc i} data can be influenced by several systematic effects. For {such} data, tests have shown that there is a minimum number of independent points necessary to get a reasonable description of a rotation curve (see Chapter 3.4 of \citealt{Bosma1978}\footnote{http://ned.ipac.caltech.edu/level5/March05/Bosma/frames.html}). For highly inclined galaxies, radial velocities based on moment equations are skewing the data to velocities closer to the systemic velocity (e.g. \citealt{Sancisi1979}), and the envelope tracing method advocated by \citet{Sofue1996,Sofue1997} might be more appropriate. {One possible solution is} supplementing the H{\sc i} data with higher resolution data in the inner parts using either H$\alpha$ or CO studies (e.g. \citealt{vanderKruitBosma1978}; \citealt{CorradiBosma1991}; \citealt{Sofue1997}; \citealt{Blais-Ouellette2004}; \citealt{Noordermeer2005}; LFV13). {Also, several techniques have been developed to take beam-smearing effects in account: position-velocity diagrams along the major axis (e.g., \citealt{Begeman1987}; \citealt{Broeils1992}; \citealt{Verheijen2001b}) or 3D model-cubes that consider many observational effects such as the spatial and spectral resolution, velocity dispersion or gas distribution (e.g., \citealt{Gentile2004}; \citealt{Swaters2009} and \citealt{Lelli2010,Lelli2012a,Lelli2012b,Lelli2014b}).}


In order to obtain the highest possible angular resolution for our sample of nearby galaxies, we used H$\alpha$ FP data from the observational survey introduced in \citet{Erroz-Ferrer2012} (hereafter Paper~I) and described in detail in \citealt{Erroz-Ferrer2015} (hereafter Paper~II). These data have high angular (seeing limited, $\sim$1") and spectral  ($\sim$8 km s$ ^{-1} $ sampling) resolutions, higher than data from other FP studies such as those from the Virgo survey \citep{Chemin2006} or the GHASP survey (see \citealt{Epinat2008} for a complete description of the GHASP sample and data), or typical H{\sc i} data (e.g., THINGS is one of the H{\sc i} surveys with the highest angular resolution, of $\sim6"$, \citealt{Walter2008}). 

This paper is organised as follows: Section~\ref{section6.2} gives a description of the sample selection, the observations and the data reduction and data analysis. The derived results are presented in Section~\ref{section6.4} and discussed in Section~\ref{section6.5}. Section~\ref{section6} presents our conclusions.


\section{The data}
 \label{section6.2}

\subsection{Sample, observations and data reduction}

The sample is described in Paper~II and consists of 29 galaxies spread in morphological type and other characteristics such as bar presence. Information about the galaxies can be found in Table \ref{properties6}. All these galaxies are part of the \textit{Spitzer} Survey of Stellar Structure in Galaxies (S$ ^{4} $G) \citep{Sheth2010}. The S$ ^{4} $G survey has delivered 3.6 and 4.5 $ \mu $m images of more than 2350 galaxies to the scientific community\footnote{http://irsa.ipac.caltech.edu/data/SPITZER/S4G/}. All the mid-infrared (IR) images have been processed using the S$ ^{4} $G pipelines, and we use some of the products from these for our 29 galaxies (presented in Table \ref{properties6}): (i) Surface brightness profiles, from  \citet{Munoz-Mateos2015}; (ii) Bulge-to-total ratios (\textit{B/T}) derived from the 2D structural decompositions presented in \citet{Salo2015} and (iii) the stellar mass maps \citep{Querejeta2015}. In addition, we use the mid-IR morphological classifications and \textit{T}-types from \citet{Buta2015}; and bar torques ($ Q_{\rm b} $) and bar lengths from \citet{Diaz-Garcia2015} and \citet{Herrera-Endoqui2015}.

The kinematic data have been obtained with the GH$ \alpha $FaS (Galaxy H$ \alpha $ Fabry-Perot System) instrument mounted on the William Herschel Telescope (WHT) in La Palma. In this study, we use the first and second moment maps derived from the GH$ \alpha $FaS kinematic cubes (the velocity and velocity dispersion maps). The high angular resolution of GH$ \alpha $FaS allows us to derive high-resolution rotation curves ($\sim$1 arcsec sampling). We use the ROTCUR task in {\sc gipsy}, based on the tilted-ring method explained in \citet{Begeman1989}. The velocity maps and derived rotation curves have been presented in Paper~II, along with all the details about the observations and data reduction processes.

The FP data were observed along with H$ \alpha $ narrow-band images using the Auxiliary port CAMera (ACAM), also at the WHT. The H$ \alpha $ narrow-band images have two aims. Firstly, the kinematic cubes need to be flux calibrated, and the narrow-band images allow us to flux-calibrate the intensity maps from the GH$ \alpha $FaS cubes as explained in Paper~I and Paper~II. Secondly, we can compute star formation rates (SFRs) derived from the measured fluxes in these narrow-band images.

\begin{table*}
\caption{General properties of the galaxies in the sample. Notes. (1)~Mean \textit{T}-types from the two independent morphological classifications in \citet{Buta2015}. (2)~Central surface brightness obtained following the procedure by LFV13 and measured from the light profiles derived with ellipse fitting to the 3.6 $ \mu $m images \citep{Munoz-Mateos2015}. (3)~Bar strengths derived from the torque maps obtained from the 3.6 $ \mu $m images \citep{Diaz-Garcia2015}. (4) Bar lengths from \citet{Diaz-Garcia2015} and \citet{Herrera-Endoqui2015}. (5) \textit{B/T} ratios from the structural 2D decompositions to the 3.6 $ \mu $m images \citep{Salo2015}. (6) Total stellar masses ($M_{*}$) derived using the absolute magnitudes at 3.6 and 4.5 $ \mu $m (from the ellipse fitting, \citealt{Munoz-Mateos2015}), and derived using the recipes of \citet{Querejeta2015}. The uncertainties in $ M_{*} $ include those arising from the uncertainty of 0.2 dex on $M/L$  and the uncertainties in the absolute magnitudes.}
 \label{properties6}
\centering
\begin{tabular}{c|clc|clc|c|c|c|}
\hline
  Galaxy name &  $\langle$T$\rangle$ & $ \mu_{0} $ & $ Q_{\rm b} $ & Bar length & \textit{B/T} & log($M_{*} /M _{\sun}$) \\
   &  & (mag arcsec$ ^{-2} $) & & (kpc) & & \\
   & (1) & (2) & (3) & (4) & (5) & (6)  \\
\hline
  NGC  428 & 8.0 & 20.4 $\pm$ 0.2 & 0.29 $\pm$ 0.03 & 2.41 & 0.002 & 9.72  $\pm$ 0.13 \\
  NGC  691 & 2.0 & 17.6 $\pm$ 0.7 & 0.00 $\pm$ 0.00 & 0.00 & 0.170 & 10.70 $\pm$ 0.13 \\
  NGC  864 & 4.0 & 18.0 $\pm$ 0.5 & 0.47 $\pm$ 0.07 & 3.85 & 0.027 & 10.17 $\pm$ 0.13 \\
  NGC  918 & 6.0 & 19.6 $\pm$ 0.4 & 0.23 $\pm$ 0.02 & 0.85 & 0.008 & 10.09 $\pm$ 0.13 \\
  NGC 1073 & 5.5 & 19.4 $\pm$ 0.6 & 0.63 $\pm$ 0.08 & 4.14 & 0.000 & 9.95  $\pm$ 0.13 \\
  NGC 2500 & 6.5 & 20.4 $\pm$ 0.3 & 0.28 $\pm$ 0.03 & 1.56 & 0.002 & 9.39  $\pm$ 0.13 \\
  NGC 2541 & 7.5 & 21.0 $\pm$ 0.4 & 0.00 $\pm$ 0.00 & 0.00 & 0.000 & 9.41  $\pm$ 0.13 \\
  NGC 2543 & 3.0 & 16.1 $\pm$ 0.8 & 0.35 $\pm$ 0.08 & 5.67 & 0.165 & 10.40 $\pm$ 0.13 \\
  NGC 2712 & 2.5 & 16.0 $\pm$ 0.8 & 0.28 $\pm$ 0.05 & 4.00 & 0.170 & 10.41 $\pm$ 0.13 \\
  NGC 2748 & 4.0 & 18.2 $\pm$ 0.5 & 0.45 $\pm$ 0.03 & 1.66 & 0.034 & 10.30 $\pm$ 0.13 \\
  NGC 2805 & 5.0 & 19.7 $\pm$ 0.6 & 0.19 $\pm$ 0.01 & 2.13 & 0.002 & 10.32 $\pm$ 0.13 \\
  NGC 3041 & 3.5 & 18.7 $\pm$ 0.3 & 0.00 $\pm$ 0.00 & 0.00 & 0.043 & 10.40 $\pm$ 0.13 \\
  NGC 3403 & 5.0 & 19.4 $\pm$ 0.6 & 0.00 $\pm$ 0.00 & 0.00 & 0.000 & 10.10 $\pm$ 0.13 \\
  NGC 3423 & 4.5 & 19.1 $\pm$ 0.5 & 0.00 $\pm$ 0.00 & 0.00 & 0.055 & 9.69  $\pm$ 0.13 \\
  NGC 3504 & 1.0 & 14.5 $\pm$ 1.4 & 0.25 $\pm$ 0.06 & 4.59 & 0.364 & 10.43 $\pm$ 0.13 \\
  NGC 4151 & 0.0 & 15.8 $\pm$ 1.1 & 0.09 $\pm$ 0.02 & 6.38 & 0.443 & 9.91  $\pm$ 0.13 \\
  NGC 4324 &-1.0 & 15.8 $\pm$ 0.9 & 0.00 $\pm$ 0.00 & 0.00 & 0.326 & 10.66 $\pm$ 0.13 \\
  NGC 4389 & 1.0 & 18.3 $\pm$ 0.4 & 0.52 $\pm$ 0.06 & 2.81 & 0.000 & 9.82  $\pm$ 0.13 \\
  NGC 4498 & 7.0 & 19.6 $\pm$ 0.4 & 0.46 $\pm$ 0.07 & 2.26 & 0.000 & 9.72  $\pm$ 0.13 \\
  NGC 4639 & 2.0 & 16.4 $\pm$ 0.7 & 0.26 $\pm$ 0.04 & 2.09 & 0.112 & 10.28 $\pm$ 0.13 \\
  NGC 5112 & 6.5 & 20.1 $\pm$ 0.4 & 0.62 $\pm$ 0.06 & 1.66 & 0.000 & 10.11 $\pm$ 0.13 \\
  NGC 5334 & 6.0 & 20.4 $\pm$ 0.3 & 0.49 $\pm$ 0.08 & 2.36 & 0.001 & 10.38 $\pm$ 0.13 \\
  NGC 5678 & 3.0 & 17.5 $\pm$ 0.4 & 0.00 $\pm$ 0.00 & 0.00 & 0.037 & 10.74 $\pm$ 0.13 \\
  NGC 5740 & 2.0 & 16.0 $\pm$ 0.7 & 0.16 $\pm$ 0.03 & 3.05 & 0.125 & 10.43 $\pm$ 0.13 \\
  NGC 5921 & 3.0 & 16.2 $\pm$ 1.0 & 0.34 $\pm$ 0.06 & 6.57 & 0.111 & 10.39 $\pm$ 0.13 \\
  NGC 6070 & 5.0 & 17.9 $\pm$ 0.5 & 0.00 $\pm$ 0.00 & 0.00 & 0.045 & 10.71 $\pm$ 0.13 \\
  NGC 6207 & 6.5 & 19.2 $\pm$ 0.2 & 0.21 $\pm$ 0.02 & 2.75 & 0.000 & 10.06 $\pm$ 0.13 \\
  NGC 6412 & 6.0 & 18.8 $\pm$ 0.5 & 0.24 $\pm$ 0.02 & 1.08 & 0.005 & 10.11 $\pm$ 0.13 \\
  NGC 7241 & 5.5 & 18.5 $\pm$ 0.4 & 0.00 $\pm$ 0.00 & 0.00 & 0.000 & 10.28 $\pm$ 0.13 \\
    
         \hline
\end{tabular}
\end{table*}

\subsection{Rotation curves derived from the stellar mass maps}
\label{masscurves}

To understand the stellar mass distribution, we have estimated the circular velocity from the observed photometric data, in particular, the stellar mass maps from \citet{Querejeta2015}, using the mass-to-light ratio (\textit{M/L}=0.6) from \citet{Meidt2014}. {The stellar mass maps are the result of applying the Independent Component Analysis (ICA) method first introduced by \citet{Meidt2012} to the 3.6 and 4.5 $ \mu $m images of the galaxies. This method separates the light from old stars by removing the contribution due to PAH, hot dust and intermediate-age stars. As shown by \citet{Meidt2014} and confirmed by \citet{Norris2014} and \citet{Roeck2015}, the 3.6 $ \mu $m band has the advantage that, once dust has been corrected for, a uniform \textit{M/L}=0.6 converts the stellar flux into stellar mass, with an uncertainty smaller than 0.1 dex. The difference between the raw 3.6 $ \mu $m images and the stellar mass maps is typically between 10 and 40\%, but can be practically negligible (e.g., for NGC~691, NGC~2500 or  NGC~4324). The location of the non-stellar emission also varies within the galaxy (see \citealt{Querejeta2015} for a full explanation of the stellar mass maps computation).}

The radial force map obtained to compute the bar strength using the {\sc nirqb} code (\citealt{Laurikainen2002} and \citealt{Salo2010}) based on the {polar} method  \citep{Salo1999} is used to compute the stellar component of the circular velocity\footnote{We will refer to the stellar component of the circular velocity as $ v_{*} $, leaving the notation $ v_{\rm c} $ to the circular velocity derived from the H$ \alpha $ FP kinematic data, after inclination and asymmetric drift corrections are applied.}  as:

 \begin{equation}
v_{*}^{2}=r \frac{\partial\Phi_{*}}{\partial r}= r F_{r}.
\end{equation} 
where $F_{r}$ is the radial force obtained as the partial derivative of the potential $\Phi_{*}$. {The gravitational potential is calculated using the polar method: from the deprojected \textit{Spitzer} images, the even Fourier components $ I(r,\varphi) $ are calculated within a polar grid. Then, the gravitational potential is calculated by adding the fast Fourier transformation in the azimuthal direction in combination with a direct summation of the radial and vertical directions.} Three different disc scale heights ($h _{z} $) have been used to compute the potential: $h _{z} =0.05$~$R_{K20}$, $h _{z} =0.1$~$R_{K20}$ and $h _{z} =0.2$~$R_{K20}$, where $R_{K20}$ is the \textit{2MASS} 20 mag/arcsec$ ^{2} $ isophote. The value of $h _{z} =0.1$~$R _{K20} $ is statistically the closest to reality, based on the observations of edge-on galaxies carried out by \citet{deGrijs1997} and the study of \citet{Speltincx2008}. The other two disc thickness values provide a margin of uncertainty in our method. 

 This method yields a numerical solution to the Poisson equation consistent with the method described in \citet{Casertano1983}. For test purposes, \citet{Laurikainen2002} also applied the cartesian potential evaluation (e.g., \citealt{Quillen1994}) using a two-dimensional Fast Fourier Transformation as in \citet{Buta2001}, showing that the polar method behaves better than measuring noisy and weak structures due to the Fourier smoothing. We have not included in a detailed way the uncertainties arising from assuming a certain value of $M/L$. Assuming a typical uncertainty of 0.2 dex on $M/L$ (see \citealt{Querejeta2015}), we can estimate typical uncertainties of 30\% in $ M_{*} $, 15\% in $ v_{*} $ and 15\% in the {inner} slope. As typical uncertainties on the {inner} slope from other sources amount to some 20\%, the additional uncertainty due to $M/L$ cannot invalidate any of our results. A more detailed analysis of the effects of uncertainty in $M/L$ is presented in \citet{Diaz-Garcia2015}.

\begin{figure}
\begin{center}
 \includegraphics[width=84mm]{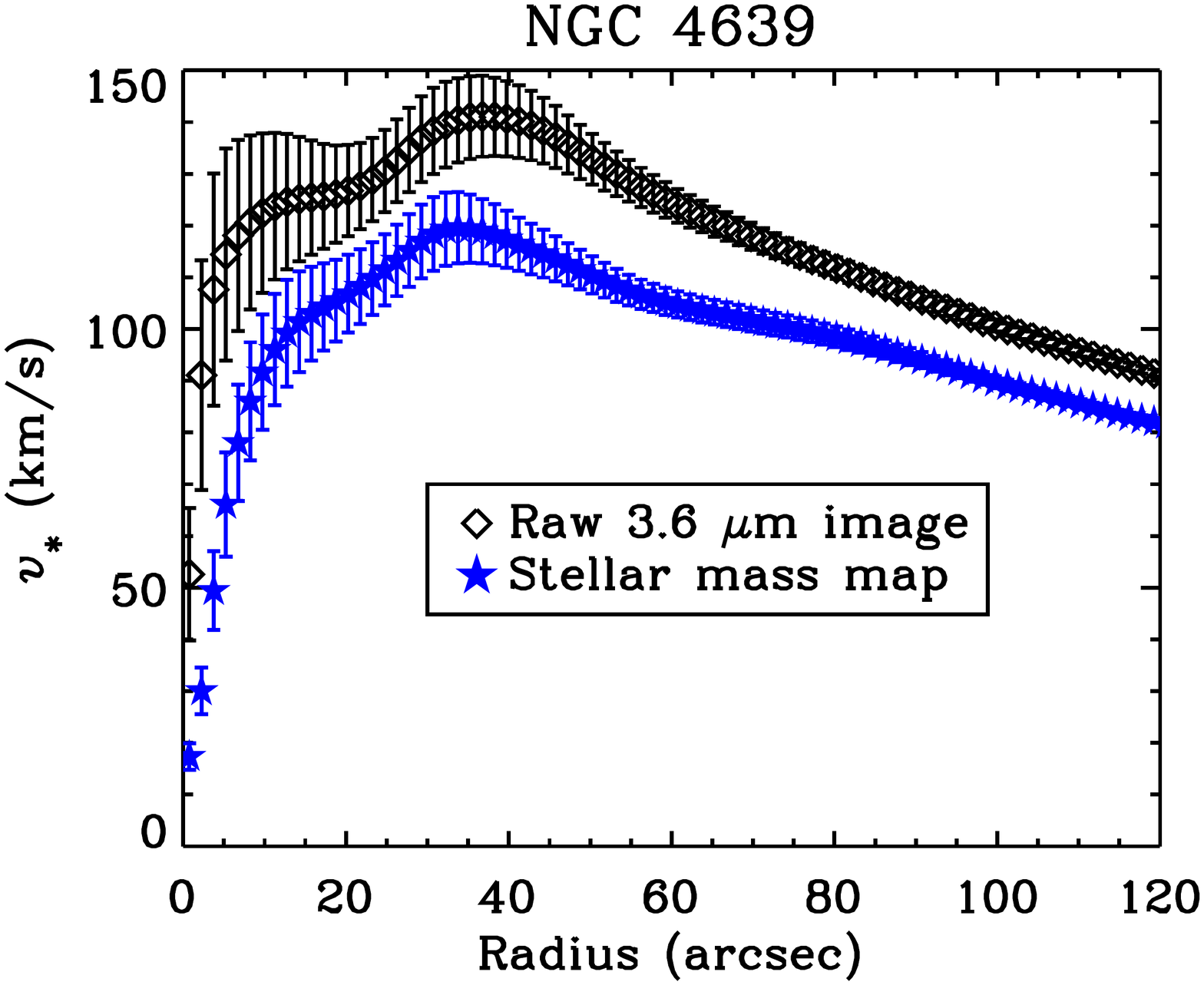}
 \includegraphics[width=84mm]{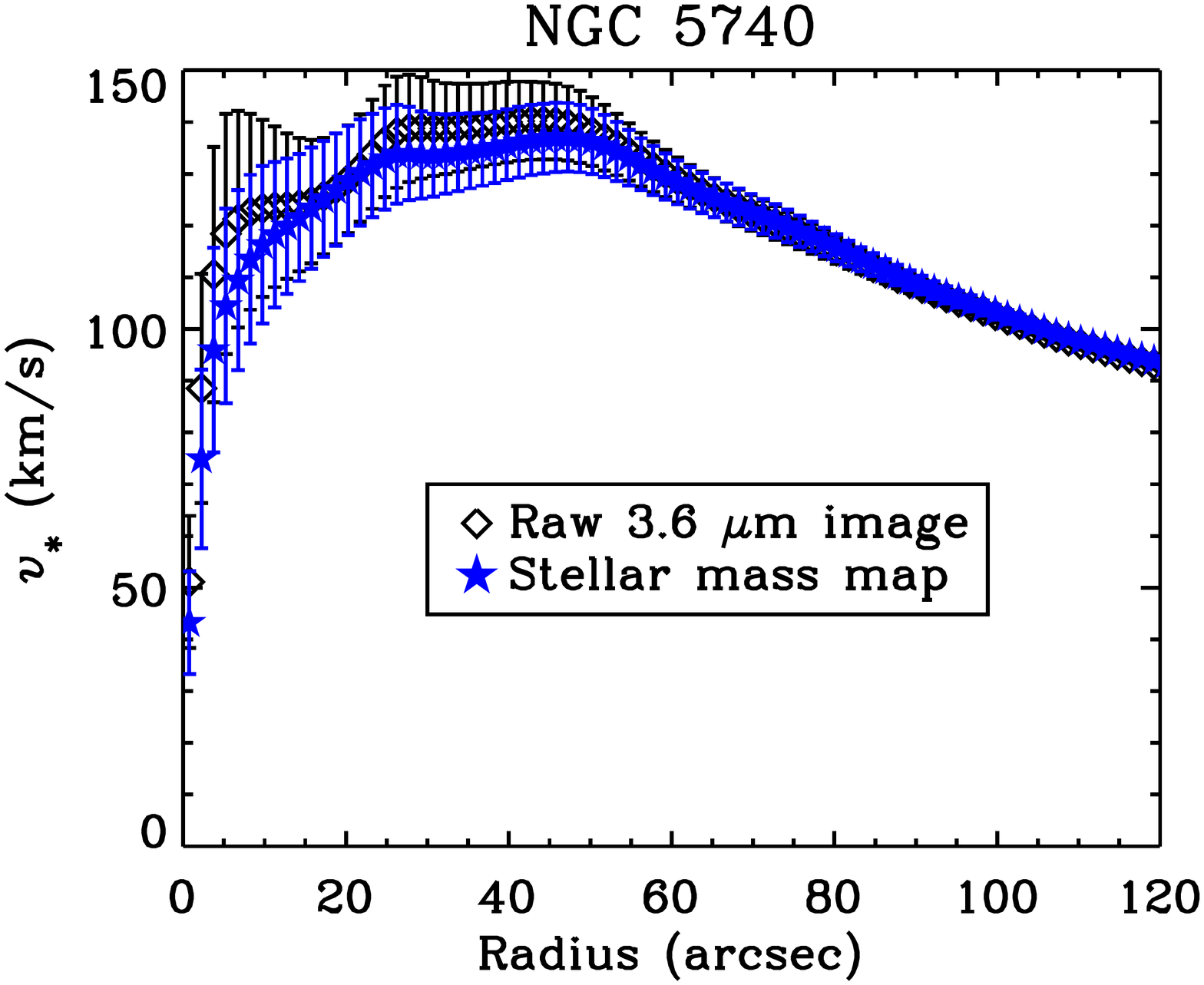}
\caption{Circular velocity curves derived from the uncorrected 3.6 $ \mu $m image (diamonds), and from the stellar mass map (blue stars) for NGC~4639 {(top) and NGC~5740 (bottom)}, assuming a scale height of $h _{z} =0.1$~$R_{K20}$ and the same $M/L$ from \citet{Meidt2014}. The uncertainties indicate the difference between assuming a scale height of $h _{z} =0.05$~$R_{K20}$ and $h _{z} =0.2$~$R_{K20}$. We have not included the uncertainties arising from assuming a certain value of $M/L$. Assuming a typical uncertainty of 0.2 dex on $M/L$ (see \citealt{Querejeta2015}), we can estimate typical uncertainties of 15\% in $ v_{*} $ and 15\% in the {inner} slope. The non-stellar emission in the raw 3.6 $ \mu $m image leads to a  significant overestimation of the derived circular velocity for NGC~4639. { For NGC~5740, the non-stellar emission in the raw 3.6 $ \mu $m image is not as important as in NGC~4639, but also leads to an overestimation of the derived circular velocity. The quantity and location of the non-stellar emission vary for each galaxy.}}
\label{4639simoncurves}
\end{center}
\end{figure}

We present in  Fig.~\ref{4639simoncurves} the circular velocity curves derived from the uncorrected 3.6 $ \mu $m images and from the stellar mass maps (hereafter stellar mass-derived rotation curves) for NGC~4639 {and NGC~5740}. We see that the difference between using the stellar mass maps and using the 3.6 $ \mu $m images is very important {for NGC~4639, but not that much for NGC~5740. In any case,} the non-stellar emission present in the 3.6 $ \mu $m images is translated into an overestimate of $v_{*}$, and the inner slope or the maximum rotational velocity is directly affected. Therefore, importantly, for this study we use the circular velocity curves ($ v_{*} $) derived from the stellar mass maps. In \citet{Diaz-Garcia2015}, we test how much $v_{*}$, gravitational torques and Fourier decompositions change when using raw 3.6 $ \mu $m data or stellar mass maps from \citet{Querejeta2015}.

\subsection{Asymmetric drift correction}

{Due to gravitational interactions between gas clouds, the observed H$\alpha$ rotation velocities can be significantly lower than the circular velocities of a test particle under the same gravitational potential. Thus, we need to correct our observed H$\alpha$ rotation curves for pressure support using the asymmetric drift correction} (ADC, see, e.g., Section 4.8.2 of \citealt{Binney2008}). The same gravitational interactions also show up in the velocity dispersion, and whenever this dispersion is significant, the ADC is not negligible. In these cases, we need to measure the velocity dispersion and compute the true circular velocity accounting for the ADC.

The circular velocity can be expressed as $ v_{\rm c}^{2}=R(\partial \Phi/\partial R) $, and following Eq. 4-227 of \citet{Binney2008}:

\begin{equation}
v_{\rm c}^{2}=v _{\phi} ^{2} + \overline{v_{\phi}^{2}} - \overline{v_{R}^{2}} - \frac{R}{\nu}  \frac{\partial (\nu \overline{v_{R}^{2}})}{\partial R}-R \frac{\partial (\overline{v_{R} v_{z}})}{\partial z}.
\end{equation}
where { $ \nu $ is the 3D density and} $v_{\phi}$, $v_{R}$ and $v_{z}$ are the azimuthal, radial and vertical components of the velocity. We set $\overline{v_{\phi}^{2}} = \sigma_{\phi}^{2}$, $ \overline{v_{R}^{2}}=\sigma_{R}^{2}$. {We assume that the scale height of the disc is constant with radius, and therefore $\partial{\rm ln}\nu/\partial{\rm ln}R=\partial{\rm ln}\Sigma/\partial{\rm ln}R$, where $\Sigma$ is the surface density profile (traced by the surface brightness profile). The H$ \alpha $ surface brightness profiles derived from the ACAM images (presented in the Appendix \ref{App:Appendixsbprof}) are fit to an exponential function $I=I_{0}e^{-R/R_{\rm exp}}$, and the resulting $R_{\rm exp}$ are collected in Table~\ref{propertiessb}. {The peaks in the surface brightness profiles come from the emission of the H{\sc ii} regions.}
 We assume that the velocity dispersion is isotropic, a reasonable hypothesis with gaseous discs. Thus, $\sigma_{R}=\sigma_{z}=\sigma_{\phi}=\sigma_{\rm obs}.$

\begin{table}
\caption{Results from the photometric analysis on the light profiles (H$ \alpha $ and 3.6 $ \mu $m surface brightness profiles) and exponential fits to the H$ \alpha $ velocity dispersion profiles. Column~1) Galaxy name. Column~2) Exponential radius $R_{\rm exp}$ of the H$ \alpha $ surface brightness profile. {Column~3) Exponential radius $R_{\rm exp,\sigma}$ derived from the velocity dispersion profiles.} Column~4) Mean ${\sigma}_{\rm grav}$ for those galaxies {without an exponential velocity dispersion profile}.}
 \label{propertiessb}
\centering
\begin{tabular}{c|c|c|c|}
\hline
  Galaxy name  & $R_{\rm exp}$  & $R_{\rm exp,\sigma}$ & $\overline{\sigma}_{\rm grav}$  \\
 & (kpc) & (kpc) & (km s$ ^{-1} $)\\
\hline
  NGC  428 & 2.97  $\pm$ 0.25&    -  & 5.03 $\pm$ 0.45 \\ 
  NGC  691 & 9.19  $\pm$ 0.54&    -  & 3.15 $\pm$ 0.34 \\ 
  NGC  864 & 2.85  $\pm$ 0.18&   2.20 $\pm$ 0.34 &  -  \\             
  NGC  918 & 3.60  $\pm$ 0.09&    -  & 4.69 $\pm$ 0.36 \\ 
  NGC 1073 & 5.30  $\pm$ 0.54&    -  & 5.06 $\pm$ 0.23 \\ 
  NGC 2500 & 2.60  $\pm$ 0.37&    -  & 4.25 $\pm$ 0.24 \\ 
  NGC 2541 & 3.45  $\pm$ 0.32&    -  & 6.12 $\pm$ 0.33 \\ 
  NGC 2543 & 3.38  $\pm$ 0.12&  2.16  $\pm$ 0.40 &  -  \\             
  NGC 2712 & 3.08  $\pm$ 0.07&  3.29  $\pm$ 0.38 &  -  \\             
  NGC 2748 & 1.43  $\pm$ 0.05&  3.06  $\pm$ 0.58 &  -  \\             
  NGC 2805 & 15.49 $\pm$ 0.60&    -  & 6.02 $\pm$ 0.33 \\ 
  NGC 3041 & 4.13  $\pm$ 0.45&    -  & 3.40 $\pm$ 0.18 \\ 
  NGC 3403 & 2.99  $\pm$ 0.10&    -  & 3.68 $\pm$ 0.20 \\ 
  NGC 3423 & 1.65  $\pm$ 0.10&    -  & 5.53 $\pm$ 0.32 \\ 
  NGC 3504 & 3.27  $\pm$ 0.10&   3.75 $\pm$ 0.60 &  -  \\             
  NGC 4151 & 5.96  $\pm$ 0.12&    -  & 4.62 $\pm$ 0.35 \\ 
  NGC 4324 & 4.05  $\pm$ 0.19&    -  & 3.39 $\pm$ 0.21 \\ 
  NGC 4389 & 2.20  $\pm$ 0.41&  3.33  $\pm$ 0.78 &  -  \\             
  NGC 4498 & 0.87  $\pm$ 0.04&  2.08  $\pm$ 0.85 &  -  \\             
  NGC 4639 & 1.72  $\pm$ 0.07&    -  & 3.98 $\pm$ 0.26 \\ 
  NGC 5112 & 3.16  $\pm$ 0.15&    -  & 4.68 $\pm$ 0.23 \\ 
  NGC 5334 & 3.51  $\pm$ 0.09&    -  & 3.30 $\pm$ 0.41 \\ 
  NGC 5678 & 3.33  $\pm$ 0.10&    -  & 5.53 $\pm$ 0.30 \\ 
  NGC 5740 & 1.83  $\pm$ 0.10&    -  & 3.71 $\pm$ 0.34 \\ 
  NGC 5921 & 5.51  $\pm$ 0.33&    -  & 3.42 $\pm$ 0.22 \\ 
  NGC 6070 & 2.99  $\pm$ 0.15&    -  & 6.22 $\pm$ 0.44 \\ 
  NGC 6207 & 1.82  $\pm$ 0.05&    -  & 6.12 $\pm$ 0.34 \\ 
  NGC 6412 & 1.97  $\pm$ 0.26&   2.33 $\pm$ 0.54 &  -  \\             
  NGC 7241 & 2.45  $\pm$ 0.07&    -  & 5.27 $\pm$ 0.44 \\ 
           \hline
\end{tabular}
\end{table}


Unlike the stellar velocity dispersion, the observed velocity dispersion in the ionised gas results from different phenomena: gravitational interaction between particles, thermal broadening, turbulent motions, natural broadening of the line  ($\sigma_{\rm N}$) and instrumental width ($\sigma_{\rm inst}$).  The only {phenomena} that reduces the  {observed} velocity is the gravitational interaction, so we need to identify the other contributions and subtract them from the observed velocity dispersion. {We assume that the turbulent and thermal components are physical drivers of the gravitational velocity dispersion ($\sigma_{\rm grav}$), and consider $\sigma_{\rm grav}$ the result of both thermal and turbulent motions in hydrodynamical equilibrium with the gravitational potential (e.g., \citealt{Westfall2011})}. The instrumental width for each galaxy was obtained from a data cube taken with a calibration lamp, following the procedures in \citet{Relano2005}, and is 8.3~km~s$ ^{-1} $. Here, we also need to subtract the natural broadening of the line ($\sigma_{\rm N} \approx 3$ km s$ ^{-1} $, \citealt{ODell1988}). We obtain:

\begin{equation} 
\sigma_{\rm obs}^{2} = \sigma_{\rm grav}^{2} + \sigma_{\rm N}^{2} + \sigma_{\rm inst}^{2}.
\end{equation}

 If we assume a S\'ersic profile \citep{Sersic1963}, the ADC-corrected circular velocity can be expressed using eq. A.6 of \citet{Lelli2014}  (hereafter LFV14):
 
\begin{equation} \label{eq3}
v_{\rm c}^{2}=v _{\rm rot} ^{2} + \sigma_{\rm grav}^{2}\left[\dfrac{b_{n}}{n}\left( \dfrac{R}{R_{\rm eff}}\right) ^{1/n}-2\dfrac{\partial{\rm ln}\sigma_{\rm grav}}{\partial{\rm ln}R}\right]
\end{equation}
where $ n $ is the S\'ersic index, $ b_{n} $ a constant dependant on $ n $ (see \citealt{Ciotti1991} and \citealt{Ciotti1999}) and $ R_{\rm eff} $ the effective radius. {For an exponential profile, $n=1$, $ b_{1} =1.678$  and $R _{\rm eff} $ = 1.678$ \times R_{exp}$.} 

We have derived azimuthally-averaged velocity dispersion profiles from the second moment maps of the galaxies (Paper II), and presented them in the Appendix~\ref{App:Appendixsigmaprof}. For NGC~864, NGC~2543, NGC~2712, NGC~2748, NGC~3504, NGC~4389, NGC~4498 and NGC~6412, it is reasonable to assume an exponential function to $\sigma_{\rm grav}$, and then eq.~\ref{eq3} simplifies to {$v_{\rm c}^{2}=v _{\rm rot} ^{2} + \sigma_{\rm grav}^{2}(R/R_{\rm exp}+2R/R_{\rm exp,\sigma})$, where $R_{\rm exp,\sigma}$ is the exponential scale length of the velocity dispersion profiles}. For the remaining galaxies, we assume a constant value of the dispersion $\overline{\sigma}_{\rm grav}$, {although it does not fully represent the data as in most of the cases, the velocity dispersion profiles are dominated by the peaks from the H{\sc ii} regions.} For these galaxies, eq.~\ref{eq3} simplifies to $v_{\rm c}^{2}=v _{\rm rot} ^{2} + \overline{\sigma}_{\rm grav}^{2}(R/R_{\rm exp})$. These values of {$R_{\rm exp,\sigma}$ and} $\overline{\sigma}_{\rm grav}$ are collected in Table~\ref{propertiessb}.}

The resulting circular velocity curves are therefore corrected for the slowdown caused by the gravitational interactions. We have implemented the ADC on our rotation curves and thus obtained the circular velocity curves. By definition, this correction is more significant in galaxies with higher $ \sigma $. Also, in Eq. \ref{eq3} we see that the higher the scalelength ($ R_{\rm exp} $), the lower the correction. Therefore, the ADC is negligible in the central regions of most of the galaxies of our sample, like NGC~3041, where the resulting $v_{\rm c}$ curve is very similar to the observed rotation curve (Fig.~\ref{rotcurngc3041}). In Appendix \ref{App:AppendixA}, we  present the rotation curves and the ADC-corrected (circular) velocity curves for all the galaxies of our sample..

\subsection{Determination of the rotation curve {inner} slopes}
\label{section6.3.4}

The motivation of this paper is to study the inner part of the rotation curve. To characterise this, we proceed as LFV13 by measuring the {inner} slope of the curve, hereafter $ d_{R}V(0) $. The method consists of fitting a polynomial function with the form $V(R)= \sum\limits_{n=1}^{m}~a_{n}\times~R^{n}$ to the inner part of the circular velocity curve (up to the radius where the curve reaches 90~\% of the maximum velocity, $ R_{90} $). The slope of the inner part is the term of the first order $a_{1}= d_{R}V(0)= \lim\limits_{R \to 0}dV/dR$, where the fit is forced to pass through $R=0$ and $V=0$. We refer the reader to LFV13 for more details about the method. 

\begin{figure}
\begin{center}
 \includegraphics[width=84mm]{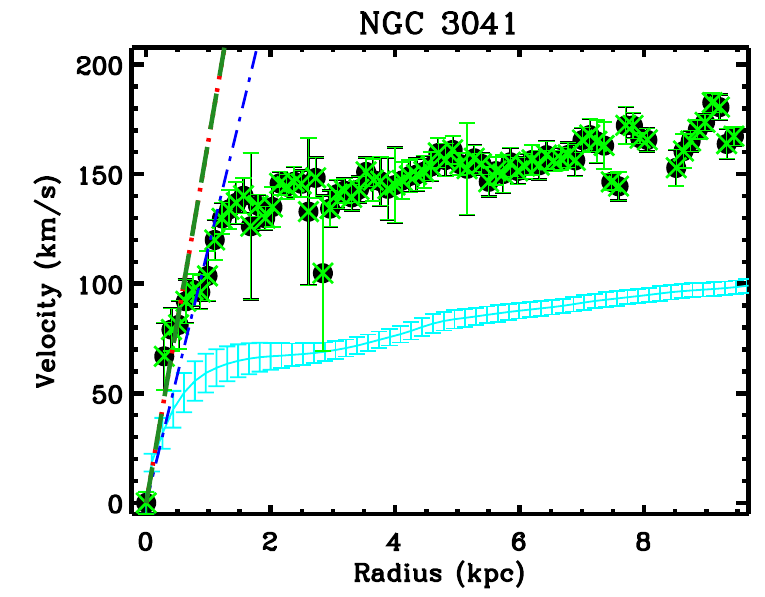}
\caption{High resolution (1") rotation curves for NGC~3041 derived from the FP data (circles), from the stellar mass maps (solid blue curve with errormarks), and after the ADC (circular velocity, green crosses). The {inner} slope of the curve has been highlighted with a red dotted line for the FP data, with a blue dot-dash-dot line for the stellar mass rotation curve and with a green dashed line for the circular velocity curve. The polynomial is fitted to the points with radius below $ R_{90} $ (the radius where the curve reaches 90~\% of the maximum velocity). The ADC is not important in this galaxy, and the circular velocity curve is very similar to the observed rotation curve. Most cases are like this one.}
\label{rotcurngc3041}
\end{center}
\end{figure}

We have measured the {inner} slope of the observed H$ \alpha $ rotation curves $d_{R}v_{\rm rot}(0)$, of the ADC-corrected circular rotation curves for the FP data $d_{R}v_{\rm c}(0) $ and of the stellar mass-derived rotation curves $ d_{R}v_{*}(0)$, and presented the results in Table~\ref{tableslopes}. In Fig \ref{rotcurngc3041}, we show the resulting fits to the observed, circular and mass rotation curves for our example galaxy NGC~3041, and the corresponding determination of the {inner} slopes, represented as the tangent line to the rotation curve at $R=0$. Again, the rotation curves and fits for all the galaxies of the sample are presented in the Appendix \ref{App:AppendixA}. 

\begin{table*}
\caption{{Results from the polynomial fits to the different rotation curves. Column I) Galaxy name. Columns II-IV) Inner} slopes of the observed H$ \alpha $ rotation curves $d_{R}v_{\rm rot}(0)$, of the ADC-corrected circular rotation curves ($d_{R}v_{\rm c}(0) $) and of the stellar mass-derived rotation curves ($ d_{R}v_{*}(0)$) correspondingly. Columns V-VII) {Radius where the curve reaches 90\% of the maximum velocity ($ R_{90} $) for the observed H$ \alpha $ rotation curves ($R_{90,\rm rot} $), for the ADC-corrected circular rotation curves ($R_{90,\rm c} $) and for the stellar mass-derived rotation curves ($R_{90,*} $) correspondingly. Columns VIII-X) Order of the polinomial fit (\textit{m}) to the observed H$ \alpha $ rotation curves ($m_{\rm rot} $), the ADC-corrected circular rotation curves ($m_{\rm c} $) and for the stellar mass-derived rotation curves ($m_{*} $) correspondingly.} \textit{Notes}) $ ^{a} $ Lower limits of $d_{R}v_{\rm c}(0) $ for galaxies whose {central parts} were not fully sampled.}
 \label{tableslopes}
\centering
\begin{tabular}{c|c|c|c|c|c|c|c|c|c}
\hline
  Galaxy name & $d_{R}v_{\rm rot}(0)$ & $d_{R}v_{\rm c}(0) $&  $ d_{R}v_{*}(0)$ &$R_{90,\rm rot} $&$R_{90,\rm c} $&$R_{90,*} $&$m_{\rm rot} $&$m_{\rm c} $&$m_{*} $ \\
   & (km s$ ^{-1} $ kpc$ ^{-1} $)& (km s$ ^{-1} $ kpc$ ^{-1} $)& (km s$ ^{-1} $ kpc$ ^{-1} $) &(kpc)&(kpc)&(kpc) \\
\hline
 NGC  428 & 65     $\pm$  22  &  65      $\pm$ 22&  64   $\pm$  2    & 5.74 & 5.74 & 2.83 & 4 & 4 & 3\\
 NGC  691 & 139    $\pm$  40  &  139     $\pm$ 40&  236  $\pm$  18   & 3.89 & 3.89 & 3.76 & 2 & 2 & 4\\
 NGC  864 & 117    $\pm$  34  &  122     $\pm$ 39&  60   $\pm$  2    & 2.48 & 2.48 & 6.16 & 2 & 1 & 4\\
 NGC  918 & 108    $\pm$  23  &  108     $\pm$ 23&  55   $\pm$  3    & 5.02 & 5.02 & 5.44 & 3 & 3 & 4\\
 NGC 1073 & 164    $\pm$  40  &  175     $\pm$ 42&  73   $\pm$  2    & 6.83 & 6.83 & 6.97 & 5 & 5 & 5\\
 NGC 2500 & 105    $\pm$  22  &  82      $\pm$ 17&  98   $\pm$  4    & 2.92 & 2.92 & 2.88 & 2 & 2 & 4\\
 NGC 2541 & 48     $\pm$  13  &  67      $\pm$ 14&  57   $\pm$  1    & 3.06 & 3.06 & 3.83 & 3 & 3 & 4\\
 NGC 2543 & 412    $\pm$  89  &  414     $\pm$ 89&  139  $\pm$  6    & 5.89 & 5.89 & 5.85 & 5 & 5 & 4\\
 NGC 2712 & 410    $\pm$  88  &  411     $\pm$ 89&  156  $\pm$  6    & 1.07 & 1.07 & 4.83 & 2 & 2 & 4\\
 NGC 2748 & 212    $\pm$  43  &  217     $\pm$ 44&  232  $\pm$  16   & 5.05 & 4.93 & 2.65 & 4 & 4 & 4\\
 NGC 2805 & 22     $\pm$  5   &  22      $\pm$  5&  54   $\pm$  2    & 18.85& 18.85& 7.83 & 4 & 4 & 4\\
 NGC 3041 & 165    $\pm$  36  &  165     $\pm$ 36&  117  $\pm$  4    & 7.01 & 7.01 & 7.21 & 4 & 4 & 5\\
 NGC 3403 & 211    $\pm$  47  &  211     $\pm$ 47&  117  $\pm$  7    & 4.81 & 4.81 & 2.40 & 5 & 5 & 3\\
 NGC 3423 & 122    $\pm$  35  &  122     $\pm$ 35&  102  $\pm$  3    & 5.23 & 5.23 & 4.42 & 3 & 3 & 5\\
 NGC 3504 & 639    $\pm$  163 &  639     $\pm$ 163&  720 $\pm$  136  & 0.61 & 0.61 & 0.51 & 2 & 2 & 1\\
 NGC 4151 & 650    $\pm$  245 &  650     $\pm$ 245&  440 $\pm$  66   & 4.90 & 4.90 & 0.51 & 1 & 1 & 1\\
 NGC 4324 & 133    $\pm$  27  &  133$^{a}\pm$  27&  944  $\pm$  65   & 1.29 & 1.29 & 1.34 & 1 & 1 & 4\\
 NGC 4389 & 80     $\pm$  30  &  83     $\pm$ 30&  83    $\pm$  3    & 4.18 & 4.18 & 2.56 & 5 & 5 & 3\\
 NGC 4498 & 46     $\pm$  10  &  47     $\pm$ 10&  102   $\pm$  5    & 3.38 & 3.38 & 2.20 & 2 & 2 & 3\\
 NGC 4639 & 210    $\pm$  46  &  210$^{a}\pm$  46&  275  $\pm$  23   & 1.04 & 1.04 & 1.47 & 1 & 1 & 4\\
 NGC 5112 & 39     $\pm$  15  &  41     $\pm$ 15&  50    $\pm$  2    & 6.22 & 6.22 & 7.27 & 5 & 5 & 4\\
 NGC 5334 & 38     $\pm$  8   &  38     $\pm$ 8&  37     $\pm$  1    & 8.15 & 8.15 & 7.30 & 2 & 2 & 3\\
 NGC 5678 & 243    $\pm$  50  &  243$^{a}\pm$  50&  258  $\pm$  19   & 1.21 & 1.21 & 3.03 & 1 & 1 & 4\\
 NGC 5740 & 97     $\pm$  20  &  97$^{a}\pm$   21&  422  $\pm$  30   & 3.34 & 3.34 & 2.06 & 1 & 1 & 4\\
 NGC 5921 & 42     $\pm$  9   &  42$^{a}\pm$   9&  224   $\pm$  10   & 5.40 & 5.40 & 4.10 & 1 & 1 & 4\\
 NGC 6070 & 105    $\pm$  27  &  132     $\pm$ 33&  80   $\pm$  5    & 7.74 & 7.74 & 5.99 & 4 & 4 & 3\\
 NGC 6207 & 75     $\pm$  17  &  76     $\pm$ 17&  152   $\pm$  16   & 5.25 & 5.25 & 1.28 & 4 & 4 & 3\\
 NGC 6412 & 111    $\pm$  35  &  127     $\pm$ 40&  58   $\pm$  4    & 3.27 & 3.27 & 4.22 & 3 & 3 & 4\\
 NGC 7241 & 32     $\pm$  7   &  33     $\pm$ 7&  102    $\pm$  6    & 4.07 & 4.07 & 2.85 & 1 & 1 & 3\\
             \hline
\end{tabular}
\end{table*}

We have measured the {inner} slope of the circular velocity curve from our FP data in all the galaxies of the sample. The central regions of the galaxies NGC~4151, NGC~4324, NGC~4639, NGC~5740 and NGC~5921 are not sampled. These galaxies are mostly early-type barred galaxies. We have analysed those data equally as in the other galaxies in the sample in order to obtain lower limit values of $d_{R}v_{\rm rot}(0)$ and $d_{R}v_{\rm c}(0) $, which are highlighted in the forthcoming correlations with red arrows. For completeness, we have computed for these lower limits the observed {inner} slopes that LFV13 predict\footnote{The empirical relation in LFV13 is based on \textit{R}-band photometry, and we therefore calculate the corresponding relationship for the 3.6 $ \mu $m by using the correspondence between the surface brightness profiles for the SINGS galaxies in the $ R $-band \citep{Munoz-Mateos2009} and 3.6 $ \mu $m \citep{Munoz-Mateos2015}: $ \mu_{R} - \mu_{3.6} = 0.128\mu_{3.6} +2.802$} from the $ \mu_{0} $, and these values are represented with red crosses in the forthcoming plots. The differences between the observed {inner} slopes and those predicted by Lelli's equation can be up to 30\% of log($d_{R}V(0) $), {which is translated into a factor of two in the inner slope. This factor is exactly what LFV13 report as a realistic uncertainty for galaxies with compact inner components (bars or bulges) and poorly-resolved rotation curves}. NGC~4151 is a Seyfert 1.0 \citep{Ho1997}, and the central surface brightness may yield an overestimate of $d_{R}V(0)$. We therefore include the measurements from \citet{Fricke1974} which cover the central region, and measure an {inner} slope of $d_{R}V(0)=650$~km~s$^{-1} $~kpc$^{-1} $, very similar to what would correspond to its morphological type (613~km~s$ ^{-1}$~kpc$^{-1}  $, see Sect.~\ref{section6.4.2.1}). The H{\sc i} rotation curve of NGC~4151 in \citet{Bosma1977} and \citet{Bosma1981} cannot provide further information on the central parts, as beyond their first measured point at 1 arcmin, the velocity curve is flat with $ v \sim$150 km s$ ^{-1} $.


\section{Results: Rotation curve inner slopes and physical properties of galaxies}
 \label{section6.4}

\subsection{Comparison of {inner} slopes}

We first present the relationships between the {inner} slopes measured using the different velocity curves: $d_{R}v_{\rm rot}(0)$ derived from the observed H$ \alpha $ rotation curves, $d_{R}v_{\rm c}(0) $ computed from the circular velocity curves after ADC correction, and $ d_{R}v_{*}(0)$ as derived from the stellar mass-maps. 

\begin{figure}
\begin{center}
 \includegraphics[width=84mm]{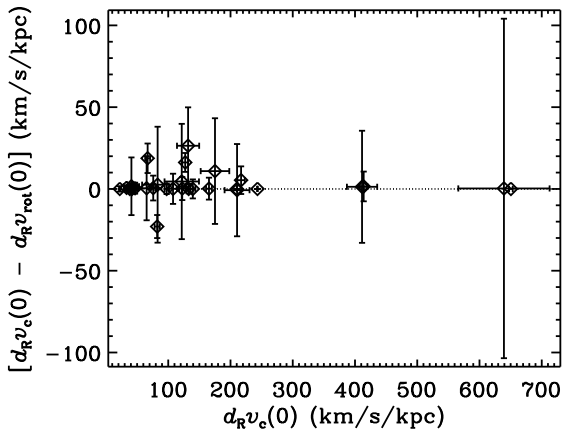}
\caption{Comparison of the {inner} slopes derived from FP rotation curve with those derived from the circular velocity (after ADC). The differences are small except for {NGC~1073, NGC~2500, NGC~2541 and NGC~6070.}}
\label{slopevcvsslopevrot}
\end{center}
\end{figure}

We see in the rotation curves (Figures in App.~\ref{App:AppendixA}) that there are very few galaxies in which ADC is important. 
{Moreover, ADC in the centre is negligible in most the cases because either the dispersion in the centre is very low or the scalelength is very large compared to the radii, leading to} $ d_{R}v_{\rm rot}(0) \approx d_{R}v_{\rm c}(0)$. In Fig.~\ref{slopevcvsslopevrot}, we compare $ d_{R}v_{\rm rot}(0) $ and $ d_{R}v_{\rm c}(0) $ and confirm that ADC { has little impact on the determination of the inner slopes.}

The circular velocity of the galaxy is defined by the gravitational potential, which takes into account the total mass of the galaxy. Therefore, we expect all the components to influence the gravitational potential of the galaxy, and therefore the derived circular velocity:

\begin{equation}
v_{\rm c}^2=v_{\rm gas}^2+v_{*}^2+v_{\rm DM}^2.
\end{equation}

\begin{figure}
\begin{center}
 \includegraphics[width=84mm]{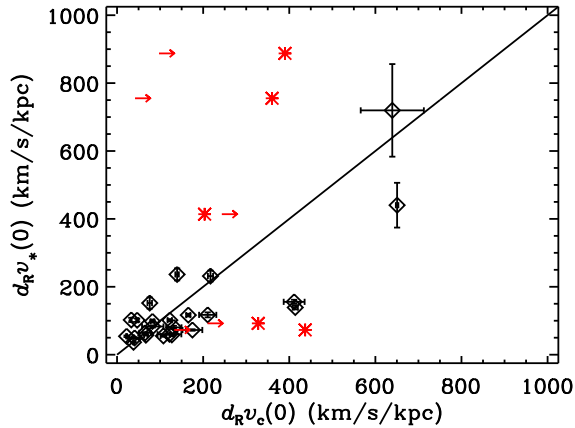}
\caption{Comparison of the {inner} slopes derived from the ADC-based $ v_{\rm c} $ with those from the stellar mass-derived rotation curves. The red arrows indicate values that are lower limits to the slopes, corresponding to galaxies with bad sampling in their central parts. We see that generally $d_{R}v_{\rm c}(0) $ $\geq$ $d_{R}v_{*}(0)$ except for the lower limits. For those cases, we have computed the {predicted} value of the {inner} slope given by the scaling relation presented in LFV13, which we represent with red crosses.}
\label{slopevsslope}
\end{center}
\end{figure}

Taking this into account, we expect to find that $v_{\rm c}^2 > v_{*}^2$, and therefore $ d_{R}v_{\rm c}(0) > d_{R}v_{*}(0)$. In Fig.~\ref{slopevsslope}, we see that the cases where $ d_{R}v_{\rm c}(0) $ $<<$ $d_{R}v_{*}(0)$ correspond to lower limits (marked with arrows) and  $ d_{R}v_{\rm c}(0) $ is probably larger than $d_{R}v_{*}(0)$. Because these galaxies have an excess of light in their centres, the rotation curve derived from the stellar mass may have been overestimated when we assume a single stellar \textit{M/L} throughout the galaxy (e.g., the starburst NGC~3504).

\subsection{Relationships with central surface brightness and \textit{T}-type}
\label{section6.4.2.1}

The aim of this study is to find possible relationships between the dynamics in the central parts of galaxies (by studying the {inner} slope of the rotation curve) and the formation, evolution and characteristics of those galaxies. First of all, we study the clearest known scaling relation between the {inner} slopes of the rotation curve and a feature of the galaxy, the one presented in LFV13. They find that $ d_{R}V(0) $ correlates with the central surface brightness of the galaxies ($ \mu_{0} $), implying that regardless of the formation or evolution of the galaxy,  the central stellar density closely relates to the inner shape of the potential well. The scaling relation for disc galaxies found by LFV13 is valid over more than two orders of magnitude in $ d_{R}V(0) $ and four orders of magnitude in $ \mu_{0} $. 

{LFV13 measure $ \mu_{0} $ differenciating between disc-dominated galaxies (for which $ \mu_{0} $ is the linear extrapolation of the luminosity profile in the inner few arcseconds to $ R $=0) and bulge-dominated galaxies (for which $ \mu_{0} $ is estimated by adding the contributions of the disc to those of the bulge, measured by extrapolating from a S\'ersic fit to the inner parts after the disc contribution has been removed).} We measure $ \mu_{0} $ from the light profiles derived with ellipse fitting to the 3.6 $ \mu $m images \citep{Munoz-Mateos2015}, following the same procedures in LFV13. {The galaxy inclinations have been obtained from Paper II.}

\begin{figure*}
\begin{center}
 \includegraphics[width=160mm]{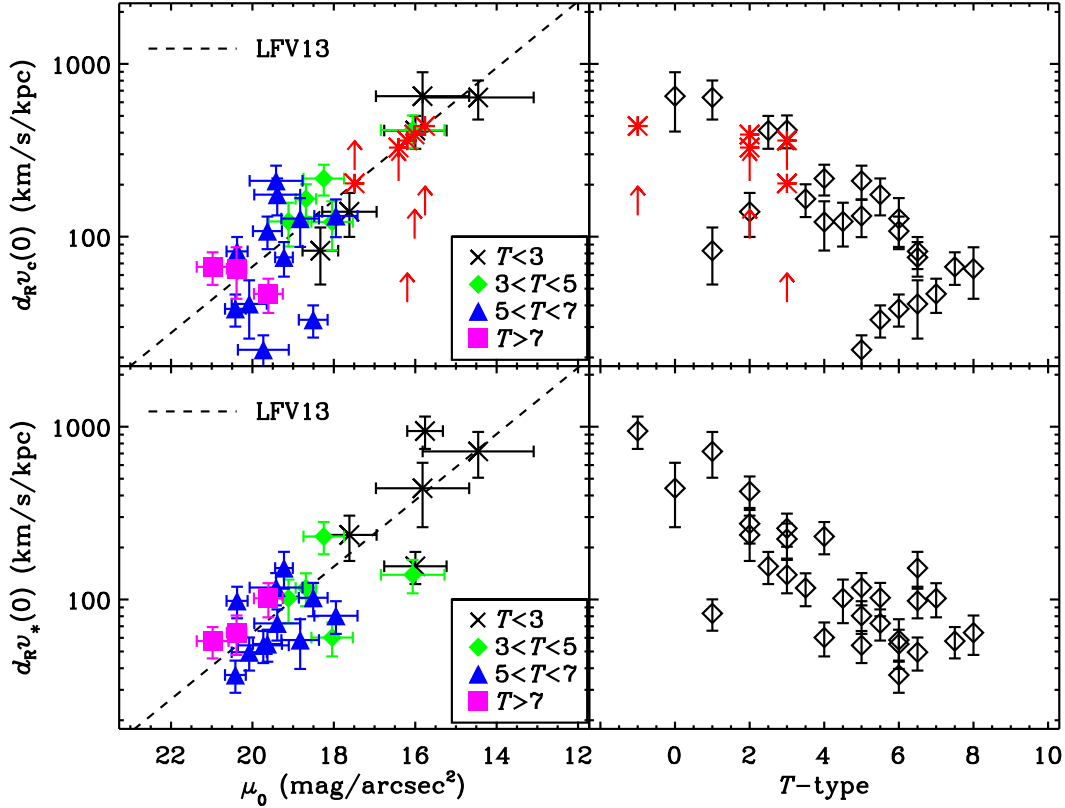}
\caption{{Inner} slopes derived from the circular velocity curves ($d_{R}v_{\rm c}(0) $, top) and from the stellar-derived curve ($d_{R}v_{*}(0)$, bottom) as a function of the central surface brightness ($ \mu_{0} $, left) and morphological \textit{T}-type (right). The dashed line corresponds to the relationship found in LFV13. As in LFV13, we have divided our sample by morphological type: diamonds correspond to galaxies with $ T $-type below 3, crosses correspond to galaxies with $3<T<5$, triangles correspond to galaxies with $5<T<7$ and squares correspond to galaxies with $T>7$. We expect the correlations between the {inner} slopes and \textit{T}-type, as the latter correlates with $ \mu_{0} $ (Fig.~\ref{typevsmu}), and $ \mu_{0} $ correlates with the logarithm of $d_{R}v_{\rm c}(0) $ (LFV13; Fig.~2). This correlation may be useful to obtain a measurement of $d_{R}v_{\rm c}(0) $ and $d_{R}v_{*}(0) $ where one does not have information on the central surface brightness but about the morphological \textit{T}-type. The red arrows correspond to lower limits of $d_{R}v_{\rm c}(0) $ (for galaxies whose {central parts} were not fully sampled). The red crosses correspond to the {predicted} {inner} slope of those lower limits (based on their $ \mu_{0} $ and following eq. 4 in LFV13).}
\label{slopevsmu}
\end{center}
\end{figure*}

We reproduce the relationship in LFV13, obtaining similar results. In Fig.~\ref{slopevsmu}, we see that $ d_{R}v_{\rm c}(0) $ and $d_{R}v_{*}(0)$ correlate with $ \mu_{0} $, with Pearson correlation coefficients of $\lvert \rho\lvert =$0.77 and $\lvert \rho \lvert=$0.83 respectively. As in the previous section, the points with the largest deviation from the {inner} slope predicted by LFV13 are the lower limits. As in LFV13, the scatter can come from observational uncertainties on the slopes, but can also be intrinsically linked to the determination of $ \mu_{0} $ with the 3.6 $ \mu $m images (affected by star formation and the different structural components such as bars, bulges or nuclear activity). The correlation with $d_{R}v_{*}(0)$ is expected, as the stellar rotation curves have been derived directly from the 3.6~$ \mu $m photometry.

\begin{figure*}
\begin{center}
 \includegraphics[width=160mm]{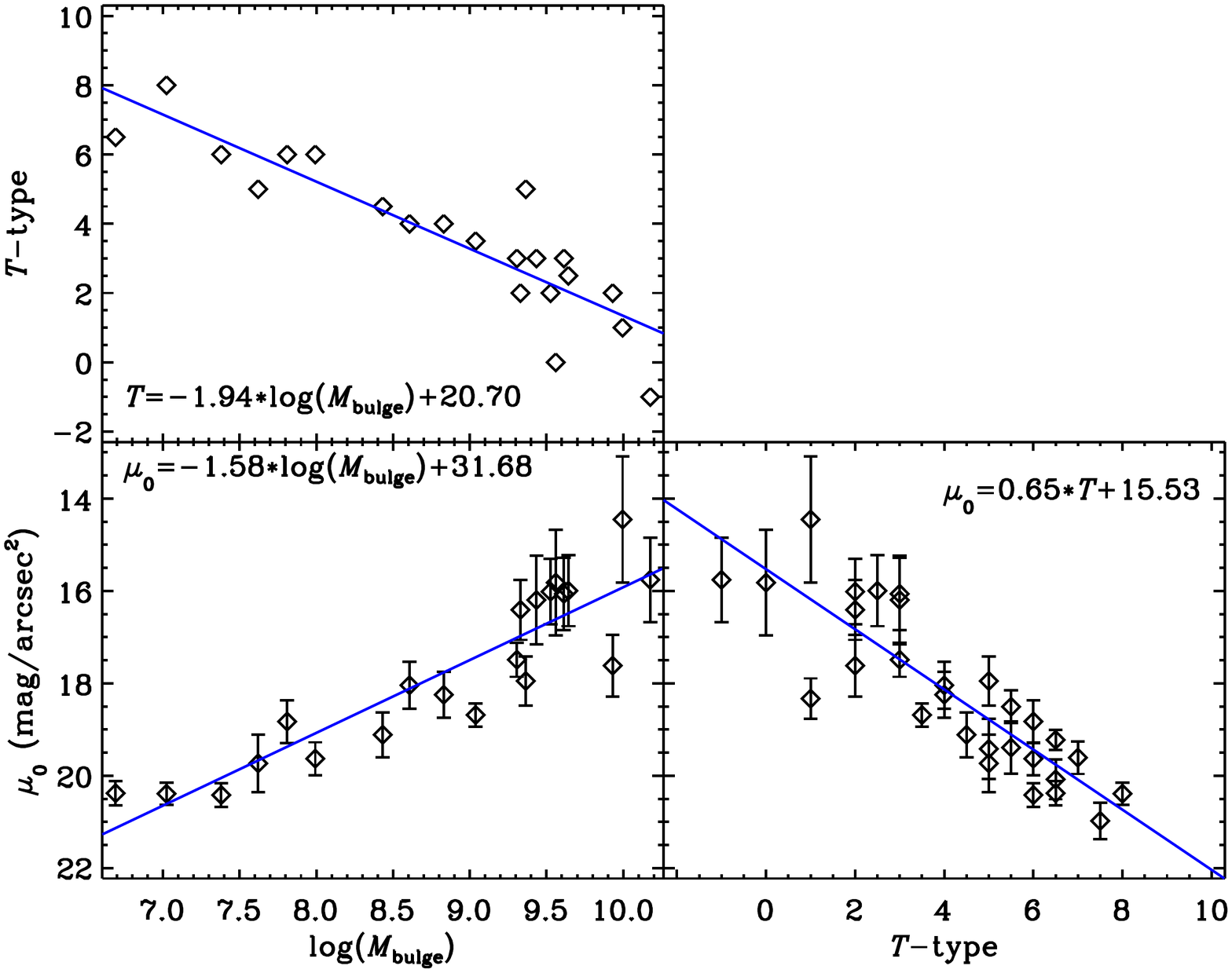}
\caption{ \textit{Left)} Correlations between the mass of the bulge ($ M_{\rm bulge} $) and the morphological \textit{T}-type (top) and $ \mu_{0} $ (bottom). \textit{Right)} Correlation between the morphological \textit{T}-type and the central surface brightness. The blue straight lines are the linear fits to the data. We see that these parameters are correlated among them.}
\label{typevsmu}
\end{center}
\end{figure*}

To better make clear the inter-relations between some galaxy properties and better understand the physical parameter that could be driving the rotation curve shape, we show in Fig.~\ref{typevsmu} the morphological \textit{T}-type and $ \mu_{0} $ as a function of the mass of the bulge ($ M_{\rm bulge} $), a parameter that directly indicates the mass in the central parts. We compute the mass of the bulge ($ M_{\rm bulge} $) as $ M_{\rm bulge} = M_{*}\times B/T$, where  $M_{*}$ is the total stellar mass and $B/T$ is the bulge-to-total ratio, assuming that the $ M/L $ does not change throughout the galaxy. {$ B/T $ is obtained from multi-component decompositions by \citet{Salo2015}, where the 2D flux-distribution is fitted with  multiple Sersic or Ferrers functions, representing the bulge, disk and bar. In some cases two disk components were fitted. In contrast to simple  bulge/disk decompositions this approach ensures that the flux of the bar is not erroneously attributed to the bulge.} $ \mu_{0} $ correlates with the morphological \textit{T}-type (Fig.~\ref{typevsmu}, top), although \textit{T} is not a physical quantity. In Fig.~\ref{typevsmu}, we show clear correlations between \textit{T}-type  and $ M_{\rm bulge} $ (top left), as well as between $ \mu_{0} $ and $ M_{\rm bulge} $ (bottom left).

Because there is a correlation between the morphological \textit{T}-type and $ \mu_{0} $ ($ \mu_{0} \approx 0.65 * T+15.53$, Fig. \ref{typevsmu}), we derive that the LFV13 relation (Eq. 8 in LFV13) would predict a correlation between the morphological \textit{T}-type and the {inner} slope of the rotation curve of ${{\rm log}[d_{R}v_{\rm c}(0)]\approx-0.14 * T}$. We find (from Fig.~\ref{typevsmu}):

\begin{align}
\label{correlationslopetype}
{\rm log}[d_{R}v_{\rm c}(0)]=-0.14(\pm 0.01)*T+2.72(\pm 0.04)~~~~{\rm and}\\
{\rm log}[d_{R}v_{*}(0)]=-0.12(\pm 0.02)*T+2.59(\pm 0.09),~~~~~~~
\end{align}


The Pearson correlation coefficient between ${\rm log}[d_{R}v_{\rm c}(0)]$ and $T$ is $\lvert \rho\lvert =0.74$, whereas that for ${\rm log}[d_{R}v_{*}(0)]$ and $T$ is  $\lvert \rho\lvert =0.79$. This relationship can also be a proxy for the computation of the {inner} slope of the inner rotation curve.  Of course, the morphological type is not a measured physical quantity  and  is defined for galaxies with and without discs, but this relationship can be useful when one does not have information about the surface brightness and wants to estimate the {inner} slope of the rotation curve.

\subsection{Relationships with stellar mass and maximum rotational velocity}

\begin{figure*}
\begin{center}
 \includegraphics[width=160mm]{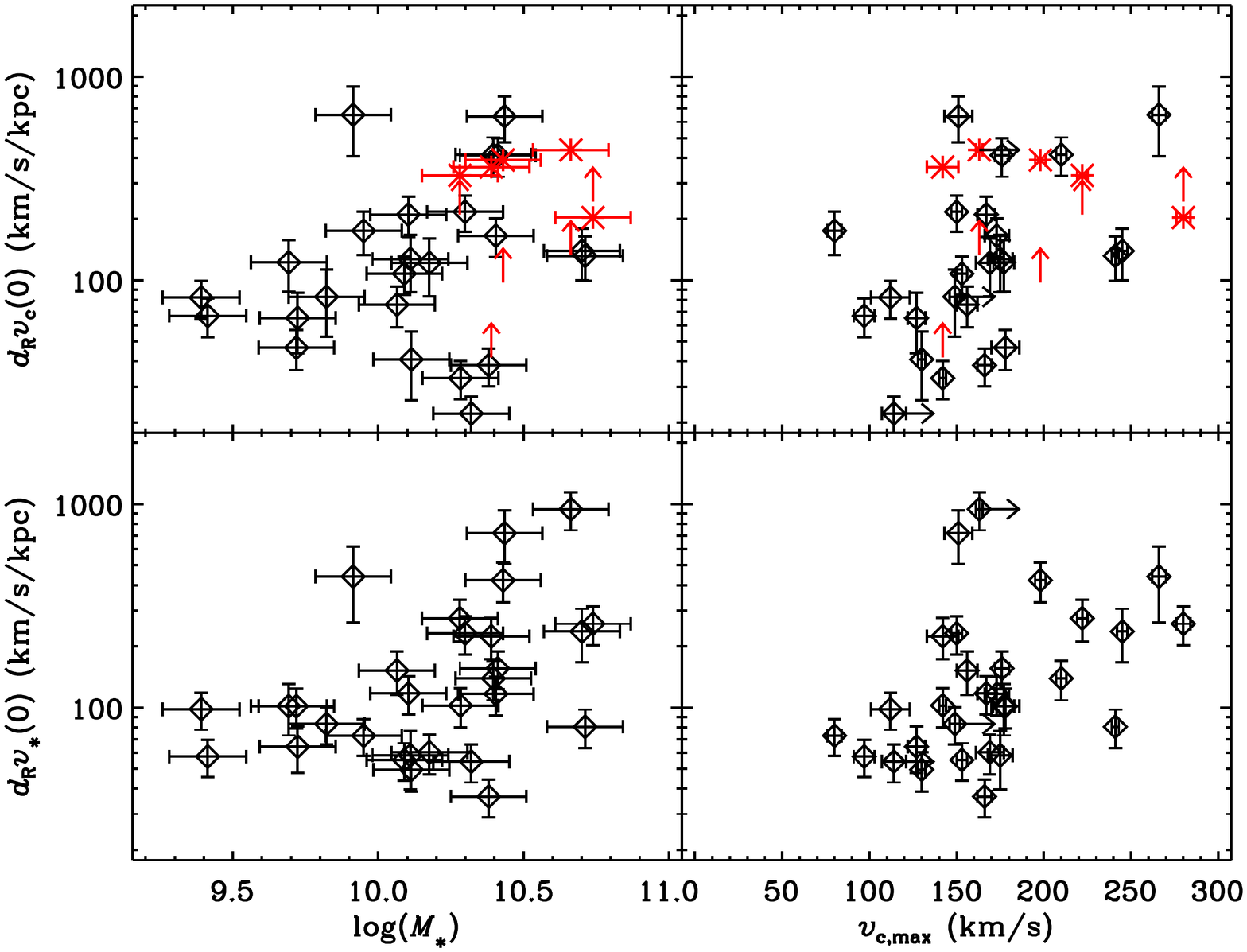}
\caption{Rotation curve {inner} slopes $ d_{R}v_{\rm c}(0) $ and $d_{R}v_{*}(0)$ as a function of the total stellar mass (log $M_{*} $, left) and the maximum circular velocity ($v_{\rm c,max} $, right). We find no correlations between the {inner} slopes and log $M_{*} $, but the total stellar mass limits the {inner} slope: steep slopes are found only in more massive galaxies, whereas lower mass galaxies present only low values of the {inner} slopes. As seen in the relationships with the total stellar mass, $v_{\rm c,max} $ limits the {inner} slopes (steeper {inner} slopes are found for galaxies with higher $v_{\rm c,max} $ and galaxies with low $v_{\rm c,max} $ just have low-rising {inner} slopes). Note that for NGC~2805, NGC~4324 and NGC~4389, $v_{\rm c,max} $ is a lower limit (marked with blue arrows that point to the right) as the curve has not reached a maximum or flat part, but continues to rise. The red arrows that point upwards correspond to lower limits of $d_{R}v_{\rm c}(0) $ (for galaxies whose {central parts} were not fully sampled). The red crosses correspond to the {predicted} {inner} slope of those lower limits (based on their $ \mu_{0} $ and following eq. 4 in LFV13).}
\label{slopevsmass}
\end{center}
\end{figure*}

In the previous Section we have seen that the central surface brightness in the inner parts of galaxies correlates with the  steepness of the inner rotation curve. We now want to know if there is also a relationship between the dynamics in the centres of galaxies and the gravitational potential of the whole galaxy. To do so, we compare the measured {inner} slopes with physical properties related with the \textit{total} mass of a galaxy: total stellar mass $M_{*} $ (which dominates the baryonic mass), and the maximum velocity. In Figure  \ref{slopevsmass} we represent $ d_{R}v_{\rm c}(0) $ and $d_{R}v_{*}(0)$ as a function of the total stellar mass (log $M_{*} $) and the maximum circular velocity ($v_{\rm c,max} $). Note that for NGC~2805, NGC~4324 and NGC~4389, $v_{\rm c,max} $ is a lower limit as we do not see that the curve has reached a maximum or flat part.

\begin{figure}
\begin{center}
 \includegraphics[width=\columnwidth]{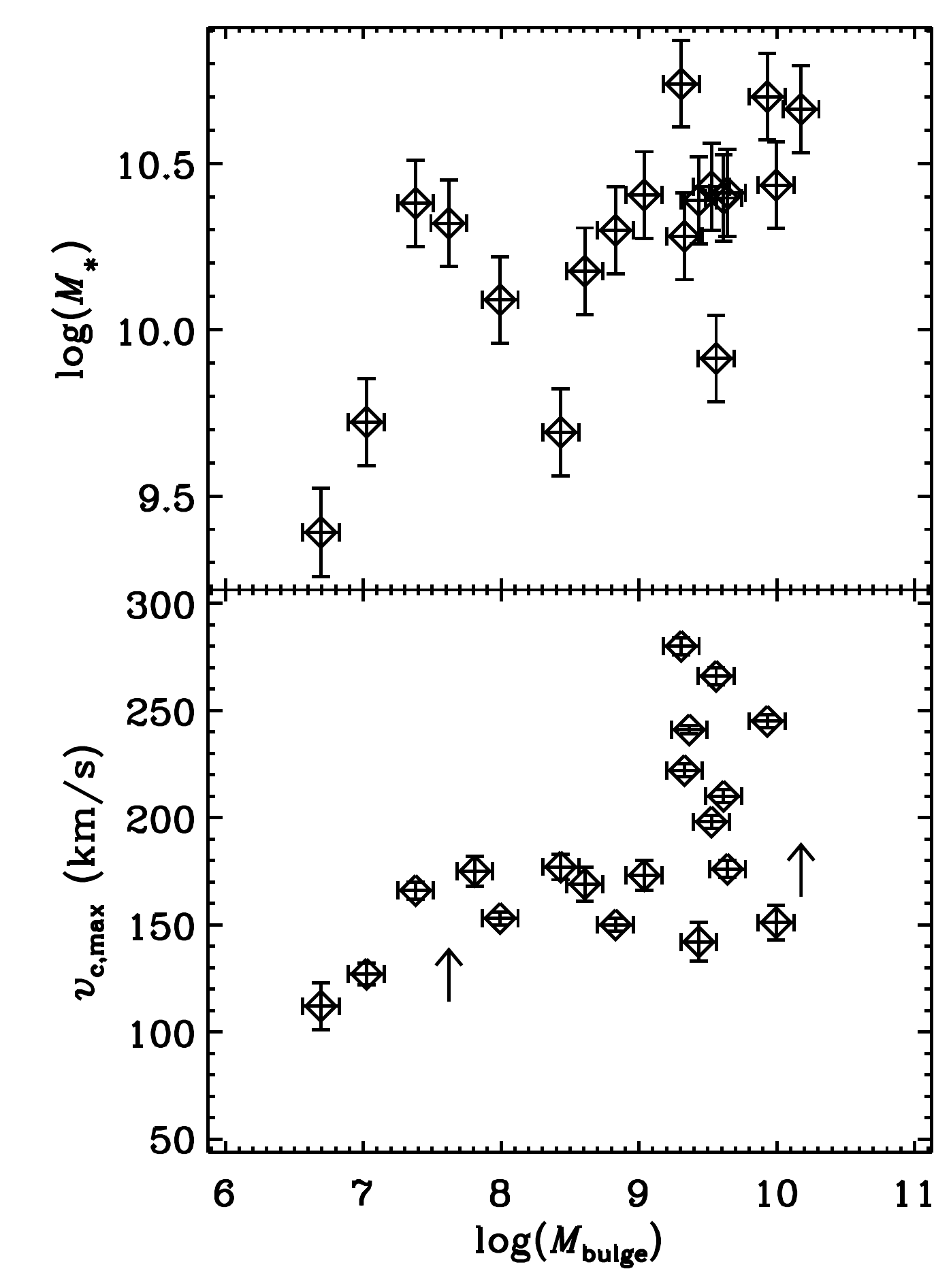}
\caption{Relationship between the total stellar mass (top) and  maximum circular velocity (bottom) as a function of the bulge mass. These parameters are well correlated, hence the weak trends between the inner slopes and $M_{*} $/$v_{\rm c,max} $ (Fig.~\ref{slopevsmass}) may be driven by the tighter correlations between $M_{\rm bulge} $ and $M_{*} $ (or $v_{\rm c,max} $), as shown here.}
\label{masstype}
\end{center}
\end{figure}

We find no linear correlations, but we can identify some trends in the sense that steeper {inner} slopes correspond to the more massive galaxies, with higher $v_{\rm c,max} $. Also, the total stellar mass and $v_{\rm c,max} $ limit the {inner} slope, and low-mass galaxies (with lower $v_{\rm c,max} $) only have slow-rising rotation curves. It is well known that for disc galaxies, the {higher the stellar mass of} a galaxy, the higher its maximum rotation velocity (Tully-Fisher relation, \citealt{Tully1977}). Here, we see that $d_{R}v_{*}(0)$ increases with $v_{\rm*,max} $. {However, these weak trends may arise from the relationship between the total stellar mass (and  maximum circular velocity) with the mass of the bulge. We see in Fig.~\ref{masstype} that there is a good correspondence between $M_{\rm bulge} $ and $M_{*} $ (or $v_{\rm c,max} $).}

\subsection{Relationships with bar and bulge parameters}
\label{section6.4.2.4}

In the previous Section, we have seen that the total stellar mass does not determine the {inner} slope of the rotation curve. Therefore, as the central surface brightness does correlate with the {inner} slope of the rotation curve (Sect. \ref{section6.4.2.1}), we need to investigate further those structural components of galaxies that play a role in galaxy evolution and participate in the distribution of light, and thus mass, within the galaxy.

\begin{figure*}
\begin{center}
 \includegraphics[width=175mm]{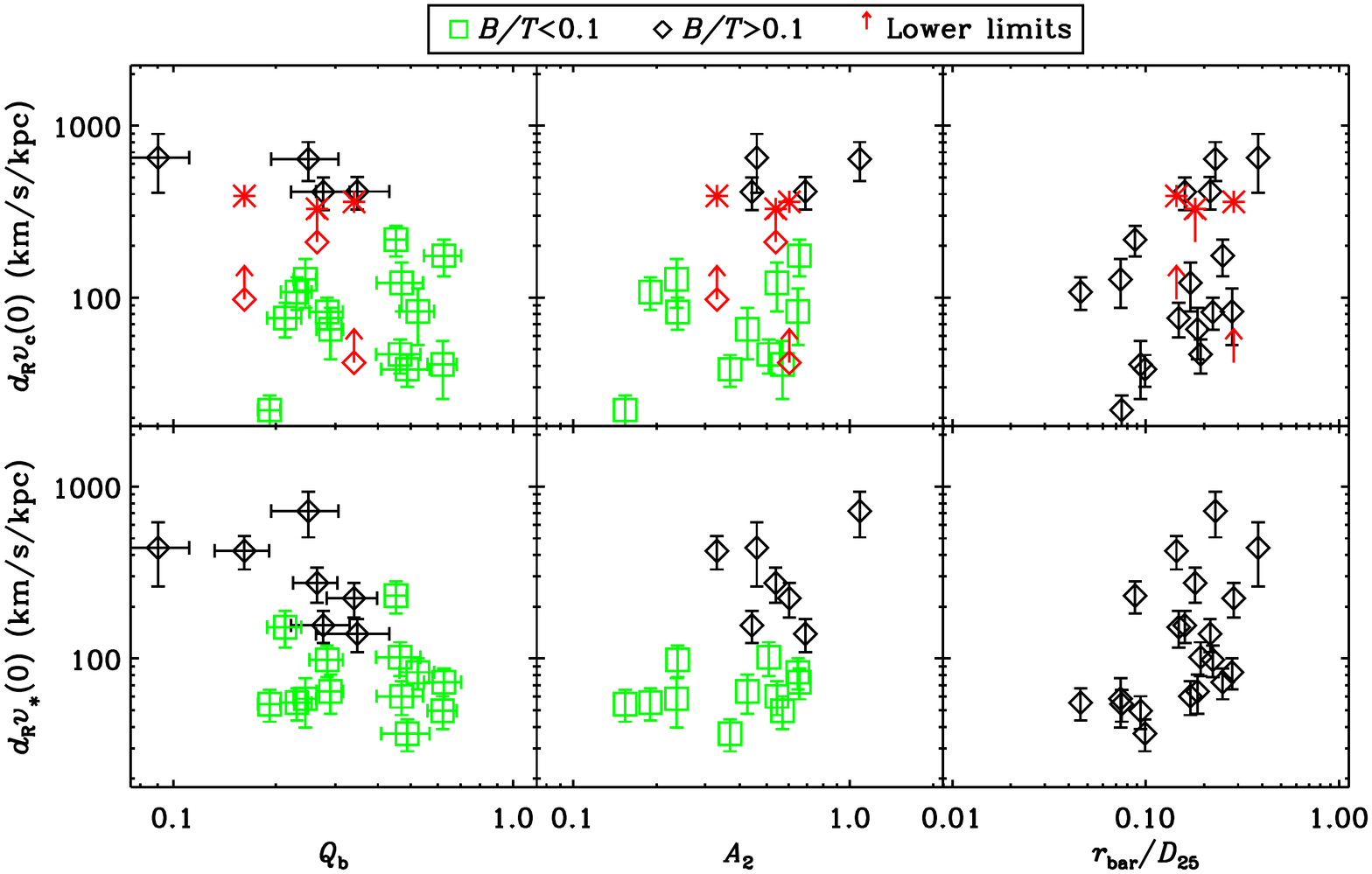}
\caption{Rotation curve {inner} slopes $d_{R}v_{\rm c}(0) $ and $d_{R}v_{*}(0)$ as a function of $ Q_{\rm b} $ (left), $ A_{2} $ (middle) and the bar length normalized by $D_{25}$ (right). Due to the possible influence of the bulge on the determination of $ Q_{\rm b} $, we have differentiated between galaxies with $B/T<0.1$ and $B/T>0.1$. All the galaxies with lower limits have $B/T>0.1$. We see that $ Q_{\rm b} $ decreases with the slope and that the higher $ A_{2} $, the steeper the slope. The red arrows correspond to lower limits of $d_{R}v_{\rm c}(0) $ (for galaxies whose {central parts} were not fully sampled). The red crosses correspond to the {predicted} {inner} slope of those lower limits (based on their $ \mu_{0} $ and following eq. 4 in LFV13).}
\label{slopevsqb}
\end{center}
\end{figure*}

\begin{figure*}
\begin{center}
 \includegraphics[width=160mm]{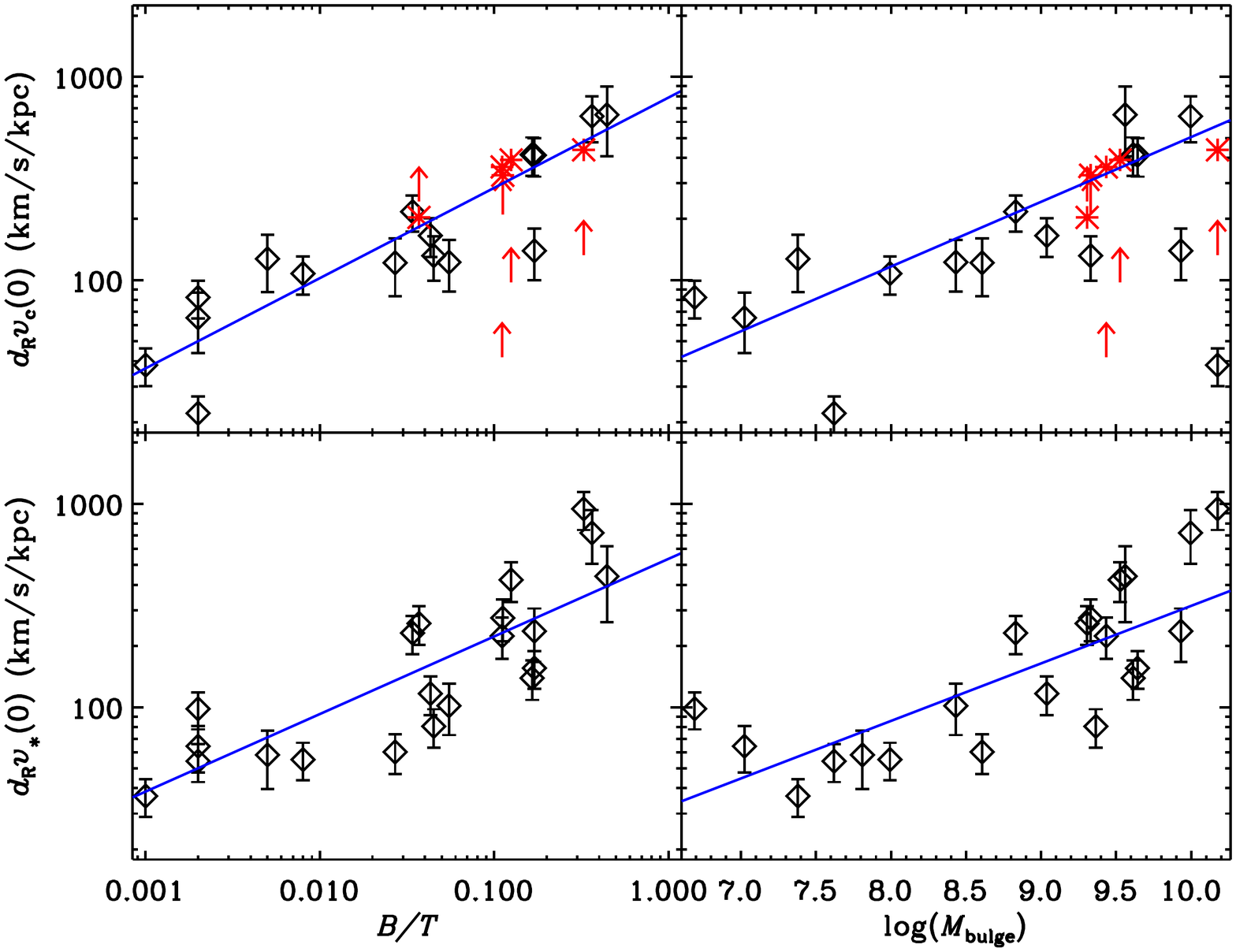}
\caption{Rotation curve {inner} slopes $d_{R}v_{\rm c}(0) $ and $d_{R}v_{*}(0)$ as a function of the bulge-to-total ratio \textit{B/T} (left) and the bulge mass $ M_{\rm bulge} $ (right). There is a moderate tendency that the more prominent the bulge, the steeper the {inner} slope. We see that the trends are clearer in the $d_{R}v_{*}(0)$ (bottom) plots. Note that only the galaxies with bulges have been considered. The red arrows correspond to lower limits of $d_{R}v_{\rm c}(0) $ (for galaxies whose {central parts} were not fully sampled). The red crosses correspond to the {predicted} {inner} slope of those lower limits (based on their $ \mu_{0} $ and following eq. 4 in LFV13).}
\label{slopevsbt}
\end{center}
\end{figure*}

We explore here whether the structural parameters (such as the presence and properties of a bar or bulge) have an impact on the shape and {inner} slope of the rotation curve. If bars and bulges modify the mass distribution of a galaxy, the dynamics should be consistent with this new distribution. To study this, we represent in Fig. \ref{slopevsqb} to \ref{slopevsbt} $ d_{R}v_{\rm c}(0) $ and $d_{R}v_{*}(0)$ as a function of bar strength (indicated by the $ Q_{\rm b} $ and $ A_{2} $ parameters), bar length and bulge-to-total light ratio (\textit{B/T}). $ Q_{\rm b} $ values have been measured from the torque maps derived from the 3.6 $ \mu $m S$ ^{4} $G images (for a description of the method see \citealt{Salo2010} and \citealt{Laurikainen2002}), and are published in a compilation of bar strengths for the S$ ^{4} $G sample \citep{Diaz-Garcia2015}. $ Q_{\rm b} $ is strongly reacting to the bulge: a stronger bulge dilutes the tangential force from the bar and lowers $ Q_{\rm b} $. Therefore, to distinguish the influence of the bulge in the bar strength, we represent galaxies with \textit{B/T} $<$ 0.1  with green squares, and galaxies with $B/T> 0.1$ with black diamonds. $ A_{2} $ is the maximum of relative Fourier (m=2) intensity amplitude $ I_{2}/I_{0} $, and has been measured from the 3.6 $ \mu $m images \citep{Diaz-Garcia2015}. $ A_{2} $ is a proxy of the strength of the bar, it measures how bright the bar is relative to the background. The radii of maximum torque in the radial force profiles have been chosen as a proxy for the bar length except when the torque maps do not present a clear butterfly pattern with an identifiable maximum at the bar region, in which case the bar lengths were measured visually \citep{Herrera-Endoqui2015}. The bar lengths have been normalised by the apparent major axis isophotal diameter (\textit{D}$ _{25} $), measured at or reduced to the surface brightness level B = 25.0 mag arcsec$ ^{-2} $, as explained in Section 3.4.a, page 21, of Volume I of \citet{RC3}. \textit{B/T} ratios have been derived from the 2D decompositions of the 3.6 $ \mu $m S$ ^{4} $G images. We refer the reader to \citet{Salo2015}, where there is information about the method and the uncertainties in the derived parameters. It is important to note that, in some cases, the excess light in 3.6 $ \mu $m due to SF (e.g., NGC~3504) as well as the presence of a bar (if it has not been included in the decompositions, e.g. NGC~4151) can lead to an overestimation of \textit{B/T}.


We find some limits, as in the relationships between {inner} slopes and mass: galaxies with high $ Q_{\rm b} $ have slowly-rising rotation curves, and steeper {inner} slopes tend to be found in galaxies with low $ Q_{\rm b} $. {Regarding $ A_{2} $}, the {inner} slopes are steeper for galaxies with higher $ A_{2} $, and slowly-rising rotation curves are found for galaxies with lower $ A_{2} $. {However, the weak trend between $ A_{2} $ and the inner slope is largely driven by the extreme values of $ A_{2} $. In Fig.~\ref{slopevsqb}, no clear correlation or trend is found when we only look at low $B/T$ galaxies. Thus, for the cases where $ Q_{\rm b} $ and $ A_{2} $ measurements are not affected by a bulge, neither  $ Q_{\rm b} $ nor $ A_{2} $ correlate with the rotation curve inner} slope. These weak trends will be discussed in Sect.~\ref{section6.5}.

When we plot $ d_{R}v_{\rm c}(0) $ and $d_{R}v_{*}(0)$ as a function of \textit{B/T} (Fig.~\ref{slopevsbt}, top), {we see a clear correlation between \textit{B/T} and the inner slopes}. Also, in the bottom panels of Fig.~\ref{slopevsbt} we see that $d_{R}v_{*}(0)$ and $ d_{R}v_{\rm c}(0) $ are correlated with the mass of the bulge, but not as much as $B/T$. In  Section \ref{section6.5}, we will discuss the physical consequences of this difference.

\subsection{Relationships with star formation}

Finally, we want to explore if the steepness of the rotation curve is influenced by the star formation. During star formation events, repeated supernova explosions in the core of the galaxies can result in not only an expulsion of gas from the central parts, but also to a re-distribution of dark matter (e.g., \citealt{Pontzen2012}). Thus, as SFR is a proxy for the supernova rate (and therefore energy injection into the potential), we might expect changes in the dynamics of the central parts of galaxies (see, e.g., LFV14). To check this, we measure the massive SFR as determined from the H$ \alpha $ images, the total SFR(H$ \alpha $). Furthermore, we measure the SFR densities ($ \Sigma $SFRs) as the SFR normalised by the physical area (in kpc$ ^{2} $).

\begin{figure*}
\begin{center}
 \includegraphics[width=160mm]{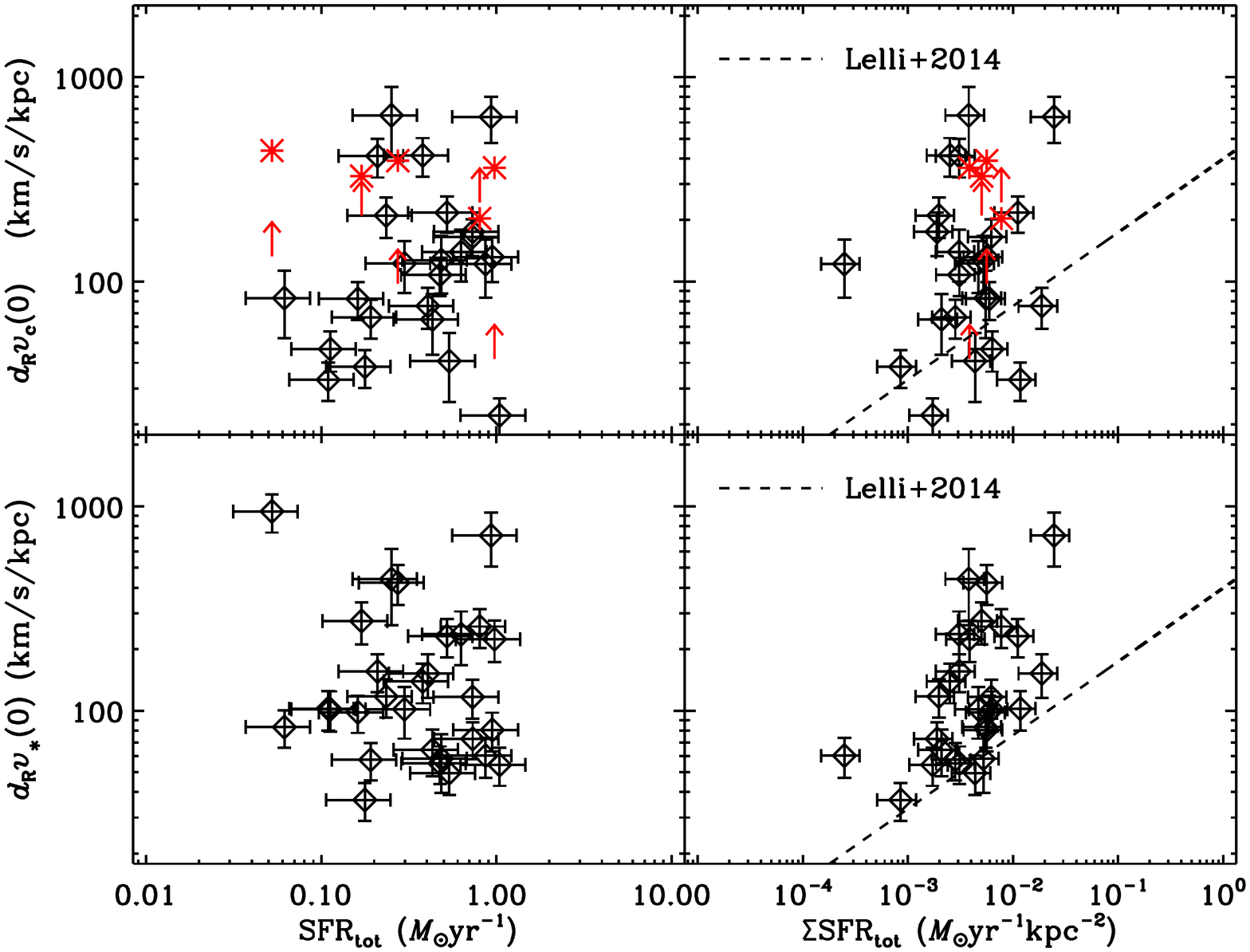}
\caption{Rotation curve {inner} slopes $ d_{R}v_{\rm c}(0) $ and $d_{R}v_{*}(0)$ as a function of the total SFR (left) and $ \Sigma $SFR$ _{\rm tot} $ (right). The red arrows correspond to lower limits of $d_{R}v_{\rm c}(0) $ (for galaxies whose {central parts} were not fully sampled). The red crosses correspond to the {predicted} {inner} slope of those lower limits (based on their $ \mu_{0} $ and following eq. 4 in LFV13). {The dashed line in the right panels shows the relation found for low-mass starburst and irregular galaxies by LFV14}.}
\label{slopevssfrtot}
\end{center}
\end{figure*}

In Fig.~\ref{slopevssfrtot} we represent $ d_{R}v_{\rm c}(0) $ and $d_{R}v_{*}(0)$ as a function of the total SFR and total $ \Sigma $SFR. There is no correlation between these quantities and the {inner} slope of the inner part of the rotation curve, although this correlation was found in LFV14 for {dwarf} starbursts and irregular galaxies (see Fig.~\ref{slopevssfrtot}).  {This lack of correlation will be discussed in Sect.~\ref{section6.5}}

%
%
%
%
%
%

\subsection{Statistical significance of linear relationships}

Our sample galaxies show a wide spread in characteristics: we have barred and unbarred galaxies, galaxies with and without bulges, and galaxies hosting AGN. {In the previous sub-sections}, we have presented several plots that relate the results from the rotation curve analysis (the slope of the inner part of the ADC-corrected circular velocity curve $d_{R}v_{\rm c}(0) $, and the slope of the inner part of the stellar mass derived velocity curve $d_{R}v_{*}(0) $) to a series of galaxy parameters. We have fitted linear functions to the {data-points} and computed the {inner} slopes as a function of these galaxy parameters. Here we collect all the resulting fits with the correspondent correlation coefficient. When we write the equation of a line as $Y = a  \times  X + b$, \textit{Y} is the dependent variable (log[$d_{R}v_{\rm c}(0) $] and log[$d_{R}v_{*}(0) $]), \textit{X} is the independent variable [$ \mu_{0} $, $ T $-type, log($ M_{*} $), $ v_{\rm max} $, log($ Q_{\rm b} $),  log($ A_{2} $), log($ r_{\rm bar}/D_{25} $), log($ B/T $), log($ M_{\rm bulge} $), log(SFR$ _{\rm tot} $) and $ \Sigma$SFR$_{\rm tot} $], and \textit{a} and \textit{b} are the {inner} slope and y-intercept of the line fitted to the points. In Table~\ref{correlations} we present \textit{a} and \textit{b} for the linear fits to both $d_{R}v_{\rm c}(0) $ and $d_{R}v_{*}(0) $. We also present the absolute value of the Pearson correlation coefficient $ \rho $ for each linear fit, where $ \lvert\rho\lvert = 0 $ means no correlation and $ \lvert\rho\lvert =1$ corresponds to a perfect linear correlation.

\begin{table*}
\caption{Linear fits and their correlation coefficients. The equation of a line is $Y = a  \times  X + b$, where \textit{Y} is the dependent variable (log[$d_{R}v_{\rm c}(0) $] and log[$d_{R}v_{*}(0) $]), \textit{X} is the independent variable (the parameters), and \textit{a} and \textit{b} are the slope and y-intercept of the line fitted to the points. We present the absolute value of the Pearson correlation coefficient $ \rho $ for each linear fit, where $ \lvert\rho\lvert = 0 $ means no correlation and $ \lvert\rho\lvert =1$ corresponds to a perfect linear correlation. The table is ordered by the correlation coefficient $ \lvert\rho\lvert$, being higher for the top four parameters (good correlation) and lower for the other parameters (not that good or lack of correlation).}
 \label{correlations}
\centering
\begin{tabular}{|c|clc|c|clc|c|c|}
  & \multicolumn{3}{|c|}{log[$d_{R}v_{\rm c}(0) ]= a  \times $ parameter + \textit{b}} & & \multicolumn{3}{|c|}{log[$d_{R}v_{*}(0) ]= a \times $ parameter + \textit{b}}  \\ \cline{2-4} \cline{6-8}  
  & \textit{a} & \textit{b} & $ \lvert\rho\lvert $& & \textit{a} & \textit{b} & $ \lvert\rho\lvert $ \\
\hline
log($ B/T $)                & 0.44  $\pm$ 0.03 & 2.90 $\pm$ 0.05 & 0.90 & & 0.38  $\pm$ 0.03 &2.73 $\pm$ 0.05 & 0.82 \\
$ \mu_{0} $                 & -0.24 $\pm$ 0.02 & 6.49 $\pm$ 0.41 & 0.84 & & -0.20 $\pm$ 0.02 &5.82 $\pm$ 0.36 & 0.83 \\
log($M_{\rm bulge}$)        & 0.32  $\pm$ 0.02 & -0.48$\pm$ 0.16 & 0.81 & & 0.28  $\pm$ 0.02 &-0.33$\pm$ 0.19 & 0.78 \\
$ T$-type                   & -0.14 $\pm$ 0.01 & 2.72 $\pm$ 0.04 & 0.74 & & -0.12 $\pm$ 0.02 &2.59 $\pm$ 0.09 & 0.79 \\
         \hline
log($A_{2}$)       & 1.16  $\pm$ 0.11 & 2.62 $\pm$ 0.05 & 0.48 & & 0.58  $\pm$ 0.11 &2.23 $\pm$ 0.05 & 0.44 \\
log($ r_{\rm bar}/D_{25} $) & 0.86  $\pm$ 0.10 & 2.87 $\pm$ 0.09 & 0.47 & & 0.58  $\pm$ 0.10 &2.52 $\pm$ 0.08 & 0.49 \\
$ v_{\rm c.max} $           & 0.004 $\pm$ 0.001& 1.45 $\pm$ 0.08 & 0.44 & & 0.003 $\pm$ 0.001&1.53 $\pm$ 0.07 & 0.43 \\
log($ Q_{\rm b} $)          & -3.29 $\pm$ 9.45 & 0.58 $\pm$ 4.87 & 0.38 & & -2.44 $\pm$ 6.91 &0.87 $\pm$ 3.57 & 0.51 \\
log($ M_{*} $)              & 0.43  $\pm$ 0.05 & -2.27$\pm$ 0.54 & 0.36 & & 0.43  $\pm$ 0.05 &-2.27$\pm$ 0.52 & 0.42 \\
log($ \Sigma $SFR)          & 0.20  $\pm$ 0.05 & 2.60 $\pm$ 0.13 & 0.18 & & 0.42  $\pm$ 0.05 &3.06 $\pm$ 0.12 & 0.53 \\
log(SFR)                    & 10.03  $\pm$ 10.15 & 6.81 $\pm$ 4.72 & 0.07 & & -5.48 $\pm$ 3.37 &-0.44 $\pm$ 1.63 & 0.12 \\
         \hline
\end{tabular}
\end{table*}

Inspection of Table~\ref{correlations} gives us hints about which parameters are important in determining the {inner} slope of the rotation curve. In the following section, we discuss the implications of the presence (or absence) of correlations as extracted from this table.


\section{Discussion}
 \label{section6.5}


{We want to explore whether we can identify changes in the distribution of mass to constrain the evolutionary processes that a galaxy has undergone. Here we study the imprints of the internal secular evolution on the internal dynamics of a galaxy. Specifically, we consider the impact of the presence of a bar, the growth of a pseudo-bulge or star formation on the measurements of the {inner} slopes of the ADC-corrected circular velocity ($ d_{R}v_{\rm c}(0) $) and of the stellar mass derived rotation curve ($d_{R}v_{*}(0)$).}

\subsection{Bars}

Bars are the most important drivers of secular evolution in galactic discs. They get stronger, longer and thinner over time (\citealt{Athanassoula2003}; \citealt{Gadotti2011}), and change the distribution of material in a galaxy, pushing the mass inside the corotation radius (CR) towards the centre, and the material outside the CR outwards.

We have in particular studied  $ Q_{\rm b} $, a parameter which represents the bar strength. By definition, $ Q_{\rm b} $ is influenced by the presence of the bulge, as $ Q_{\rm b} $ decreases with increasing axisymmetric radial force. Therefore, $ Q_{\rm b} $ anticorrelates with \textit{B/T} by definition and it is known that $ Q_{\rm b} $ increases towards later Hubble types (e.g.; \citealt{Buta2005}; \citealt{Laurikainen2007}).

In Fig. \ref{slopevsqb}, we see that $ Q_{\rm b} $ and the normalised length of the bar do not determine the {inner} slope of the rotation curve, although there is a moderate trend that the higher $ Q_{\rm b} $, the lower the {inner} slope. However, the presence of the bulge (as implicitly present in the parameter $ Q_{\rm b} $) can be seen: strong bulges dilute the tangential force. The possible anticorrelation between the {inner} slopes and $ Q_{\rm b} $ may come from the bulge dilution effect (discussed for example in \citealt{Laurikainen2004}): $ Q_{\rm b} $ is smaller for those galaxies which have steeper inner rotation curves, which means more mass concentration in the form of bulges, whereas $ Q_{\rm b} $ is larger for galaxies which have smaller bulges or no  bulges at all, but also shallower rotation curves in the central regions (for lack of a central mass concentration).

A similar scenario occurs when we use $ A_{2} $, because $ A_{2} $ increases with the presence of a bulge ($ B/T $), and the {inner} slopes increase with $ A_{2} $. We cannot prove that the strength of the bar does not influence the {inner} slope of the rotation curve, because the measurements we use depend on the presence of the bulge. We do find however that for low $ B/T $ galaxies, the bar strength as measured by $ Q_{\rm b} $ or $A_{2}$ does not significantly correlate with the rotation curve {inner} slope, suggesting that bars are not a primary driver of the {inner} slope. Bars can, however, cause local deviations from the rotational motion, what we know as non-circular motions (see Paper II). A possible caveat in this study is that the {inner} slopes have been measured from the rotation curves, which average the velocities azimuthally. Therefore, the \textit{local} deviations in the {inner} slope caused (azimuthally) by the bar may disappear in a rotation curve.

\subsection{Bulges}

Following Newton's law, we expect that the higher {the} concentration of material in the centres of galaxies, the steeper the {inner} slope. \citet{Corradi1990}, \citet{Verheijen1997}, \citet{Sancisi2004}, \citet{Noordermeer2007} and \citet{Noordermeer2008b} agree that a greater concentration of material and light (specifically, in systems with bulges) is translated into a steeper {inner} slope. Analysing the possible influence of the bulge on the {inner} slope of the rotation curve, we see in Fig.~\ref{slopevsbt} that the presence of the bulge indeed matters, as when \textit{B/T} and $ M_{\rm bulge} $ increase, $ d_{R}v_{\rm c}(0) $ and $d_{R}v_{*}(0)$ increase too, with a linear correlation with $\lvert\rho\lvert>0.75$. Also, the correlation between $ M_{\rm bulge} $ and $d_{R}v_{*}(0)$ indicates that the bulge mass plays a fundamental role in the stellar dynamics ($\lvert\rho\lvert=0.78$).

Regarding the nature of the bulges, it is possible to classify them in three types: classical bulges, disky (pseudo)bulges and boxy-peanut pseudobulges. Classical bulges are similar to elliptical galaxies (\citealt{Davies1983}; \citealt{Franx1993}; \citealt{Wyse1997} and references therein), composed by {old stellar populations} and probably formed by hierarchical merging of smaller objects. Furthermore, the redistribution of material caused by the bar torque can result in central regions with higher density, often called disky pseudobulges or simply disky bulges, which are thought to be formed through internal, secular evolution (\citealt{Athanassoula1992b}; \citealt{Wada1992,Wada1995}; \citealt{Friedli1993}; \citealt{Heller1994}; \citealt{Knapen1995}; \citealt{Sakamoto1999}; \citealt{Sheth2003}; \citealt{Regan2004}; \citealt{Kormendy2004}; \citealt{Heller2007a,Heller2007b}; \citealt{Athanassoula2008IAU}; \citealt{Fisher2009}; \citealt{Kormendy2013}). Boxy-peanut bulges are the inner parts of bars seen edge-on, see \citet{Athanassoula2005}, and might be visible in face-on view as a \textit{barlens} (\citealt{Laurikainen2011,Laurikainen2014}; \citealt{Athanassoula2014}; \citealt{Laurikainen2015}). 


From the 2D decompositions of the 3.6$ \mu $m images \citep{Salo2015}, we have derived the brightness profiles of the bulges. These follow S\'ersic (1968) function brightness profiles, which can be used as a proxy to determine if a bulge is a classical or a disky bulge (see details in, e.g., \citealt{Kormendy2004}; \citealt{Fisher2008}; \citealt{Gadotti2009}; \citealt{Laurikainen2015}). Most disky bulges have S\'ersic index  $n<2 $ (although some disky bulges have been found to have $n\approx4$), whereas many classical bulges have  $n\gtrsim2$. All the bulges in our sample have $n<2$ except for those in NGC~691, NGC~4324 and NGC~4639. Another criterion of whether a bulge is classical or disky is \textit{B/T}. Although small \textit{B/T} values do not imply that a bulge is disky, large $\textit{B/T} > 0.6$ correspond exclusively to classical bulges. All the \textit{B/T} values of our galaxies are below 0.6, so this criterion cannot confirm anything. \citet{Buta2015} classify NGC~2543 and NGC~4639 as barlenses (therefore, boxy pseudobulges). More data would be necessary to confirm the nature of our bulges, but it is likely that none of our galaxies host a massive classical bulge. However, and regardless of the nature of the bulge, the greater concentration of material in the bulge significantly influences the central dynamics of the galaxies in our sample. We see that the tendency of higher central light concentration and steeper {inner} slopes indicates that the luminous matter dominates the gravitational potential in the central parts of our galaxies. This statement is in agreement with previous studies, e.g., \citet{Lelli2014galaxies}. 

\subsection{Star formation}

Another fundamental process taking place within galaxies is SF. To understand whether SF changes the rotation curve {inner} slopes and thus the central dynamics of the galaxy, we represented the total SFR and $ \Sigma $SFRs as function of $ d_{R}v_{\rm c}(0) $ and $d_{R}v_{*}(0)$ (Fig. \ref{slopevssfrtot}). We see that there is no correlation among these parameters, implying that SF does not determine the {inner} slope of the rotation curve  {in our sample galaxies. LFV14 find a correlation between $ \Sigma $SFR and the circular velocity gradient for their sample of dwarf starbursts and irregulars. Thus, a possible explanation for the lack of correlation in this study is the difference in galaxy types with respect to those of LFV14. They study systems with one main structural component: a star-forming exponential disc, whereas our galaxies are formed by more than one structural component (i.e., bulges, spiral arms, bars). In a single-component system, the relation between internal dynamics (disc stability) and SF can be easily discerned, i.e., galaxies with steeper rotation curves have more stable gas discs, hence the gas densities can reach higher values, leading to higher SFRs. For multi-component systems, however, the central non-star-forming components will also affect the value of the inner slope of the rotation curve, and therefore a simple relation between internal dynamics and the overall star formation may be more difficult to observe.}

%
%
%

\section{Conclusions}

 \label{section6}

In this paper we have studied the relationship between the kinematics in the innermost parts of galaxies and several key galaxy parameters. We study the innermost parts (few kpc) of the rotation curves of 29 spiral galaxies of all types with high angular (seeing limited, $\sim$1") and spectral ($\sim$8 km s$ ^{-1} $ sampling) resolution FP data, and quantify the {inner} slope of the rotation curves with high precision. We compare the {inner} slopes obtained from the circular velocity curves (H$ \alpha $ rotation curves corrected for ADC), $ d_{R}v_{\rm c}(0) $, and those obtained from the velocity curves derived from stellar mass maps, $d_{R}v_{*}(0)$, with several galaxy parameters, and reach the following conclusions:
\begin{enumerate}
\item The total stellar mass and the maximum rotational velocity limit the {inner} slope: steep slopes are only found in more massive galaxies (with higher $v_{\rm c,max} $), and low-mass galaxies (galaxies with low $v_{\rm c,max} $) only have shallow {inner} slopes. {However, this trend may arise from the relationship between the total stellar mass and the mass of the bulge.}
\item We confirm that the central surface brightness ($ \mu_{0} $) correlates with the {inner} slope of the rotation curve. We find another related scaling relation for disc galaxies: a relationship between the morphological \textit{T}-type and $d_{R}v(0)$. This implies that for early-type to late-type spirals, the morphological type of a galaxy is related to the dynamics in the central parts of galaxies, quite plausibly through the radial mass distribution.
\item Although a bar can be expected to affect the distribution of material in the central parts of their host galaxies, a singular measurement of the bar strength which is unaffected by the presence of the bulge remains elusive, and therefore limit any conclusion on how bars influence the rotation curve {inner} slope. We do find, however, that for low $ B/T $ galaxies, the bar strength as measured by $ Q_{\rm b} $ or $A_{2}$ does not significantly correlate with the rotation curve {inner} slope, suggesting that bars are not a primary driver of the {inner} slope.
\item The fact that higher \textit{B/T} and $ M_{\rm bulge} $ are correlated with the {inner} slope indicates that bulges play a role in the dynamics of the central parts of galaxies. A higher concentration of stellar mass in the centres of galaxies is translated into steeper {inner} slopes, confirming that baryonic mass dominates the dynamics in the inner regions.
\end{enumerate}

\section*{Acknowledgements}

We thank Mauricio Cisternas and Dimitri Gadotti for useful comments during the preparation of this manuscript. The authors thank the entire S$ ^{4} $G team for their efforts in this project. We thank the referee for his comments, which improved the quality of the paper. We acknowledge financial support to the DAGAL network from the People Programme (Marie Curie Actions) of the European Union's Seventh Framework Programme FP7/2007-2013/ under REA grant agreement number PITN-GA-2011-289313, and from the Spanish MINECO under grant numbers AYA2007-67625-CO2-O2 and AYA2013-41243-P. This work was co-funded under the Marie Curie Actions of the European Commission (FP7-COFUND). We also gratefully acknowledge support from NASA JPL/Spitzer grant RSA 1374189 provided for the S$ ^{4} $G project. JHK thanks the Astrophysical Research Institute of Liverpool John Moores University for their hospitality, and the Spanish Ministry of Education, Culture and Sports for financial support of his visit there, through grant number PR2015-00512. E.A. and A.B. thank the CNES for support. JCMM acknowledges support from the National Radio Astronomy Observatory, which is a facility of the National Science Foundation operated under cooperative agreement by Associated Universities, Inc. HS, EL and SC acknowledge the Academy of Finland for support. This research is based on observations made with the WHT operated on the island of La Palma by the Isaac Newton Group of Telescopes, in the Spanish Observatorio del Roque de Los Muchachos of the Instituto de Astrof\'isica de Canarias. We acknowledge the usage of the HyperLeda database (http://leda.univ-lyon1.fr) This research has made use of the NASA/IPAC Extragalactic Database (NED) which is operated by JPL, Caltech, under contract with NASA.




\bibliographystyle{mnras}
\bibliography{references} 


\label{lastpage}
\end{document}